\PassOptionsToPackage{table}{xcolor}
\documentclass[aps,physrev,twocolumn,groupedaddress, floatfix]{revtex4-2}

\AtBeginDocument{%
  }

\usepackage{multirow}
\usepackage{bm}
\usepackage{graphicx}
\usepackage{url}
\usepackage{subcaption}
\usepackage{subcaption}
\usepackage[percent]{overpic}
\usepackage{stmaryrd}
\usepackage{dcolumn}
\usepackage{amssymb}
\usepackage{placeins}
\usepackage[group-separator={,}]{siunitx}

\usepackage{braket}
\usepackage{amsmath}
\usepackage{afterpage}
\usepackage{xcolor}

\begin{document}

\newcommand{\innerProduct}[2]{\langle #1 \vert  #2 \rangle} 
\newcommand{\memk}[1]{\ket{\boldsymbol{#1}}}
\newcommand{\memb}[1]{\bra{\boldsymbol{#1}}}
\renewcommand{\topfraction}{.85}
\renewcommand{\bottomfraction}{.85}
\renewcommand{\floatpagefraction}{.95}
\renewcommand{\dblfloatpagefraction}{0.95}
\renewcommand{\textfraction}{0.07}

\setcounter{topnumber}{2}
\setcounter{dbltopnumber}{2}
\renewcommand{\topfraction}{0.95}
\renewcommand{\dbltopfraction}{0.95}
\renewcommand{\textfraction}{0.05}
\renewcommand{\floatpagefraction}{0.9}
\renewcommand{\dblfloatpagefraction}{0.9}

\title{Transversal Fault Tolerant Distributed Quantum Computing Operations}

\author{John Stack}
\email{Contact author: jstack@ncsu.edu}
\author{Ming Wang}%
\author{Frank Mueller}%
\affiliation{
   Department of Computer Science,
   North Carolina State University, Raleigh, North Carolina, USA.}

\begin{abstract}
Distributed architectures are a route to scalable quantum computing,
but the performance of fault-tolerant operations across noisy
inter-module links remains poorly characterized. We present
circuit-level simulations of two key distributed primitives:
transversal non-local CNOT and logical teleportation using surface and
bivariate-bicycle codes. We then simulate the use of these distributed
primitives in a major subroutine of common quantum algorithms. The
results, enabled by our scalable library
\emph{Transversal Multiple CodeBlock Simulator}, demonstrate that on appropriate devices distributed qLDPC transversal operations can outperform surface code lattice surgery and enable efficient parallel computation with lower
Bell pair consumption. Notably, we find that the non-local CNOT
achieves up to an order of magnitude lower logical error rates than
teleportation at the same code distance and noise levels. We further
show that code distances of $d \approx 11$ at physical error rate $p \sim 10^{-4}$ and $d \approx 29$ at $p \sim 10^{-3}$, with $p_{\mathrm{ebit}}=10p$, are sufficient to achieve
logical error rates below $10^{-12}$, enabling large-scale algorithms. These results
provide critical guidance for architecture and code selection in
distributed quantum computing.

\end{abstract}

\maketitle

\section{Introduction}

\begin{figure*}[t]
  \centering
  \begin{subfigure}[t]{0.49\textwidth}
    \centering
    \includegraphics[width=\linewidth]{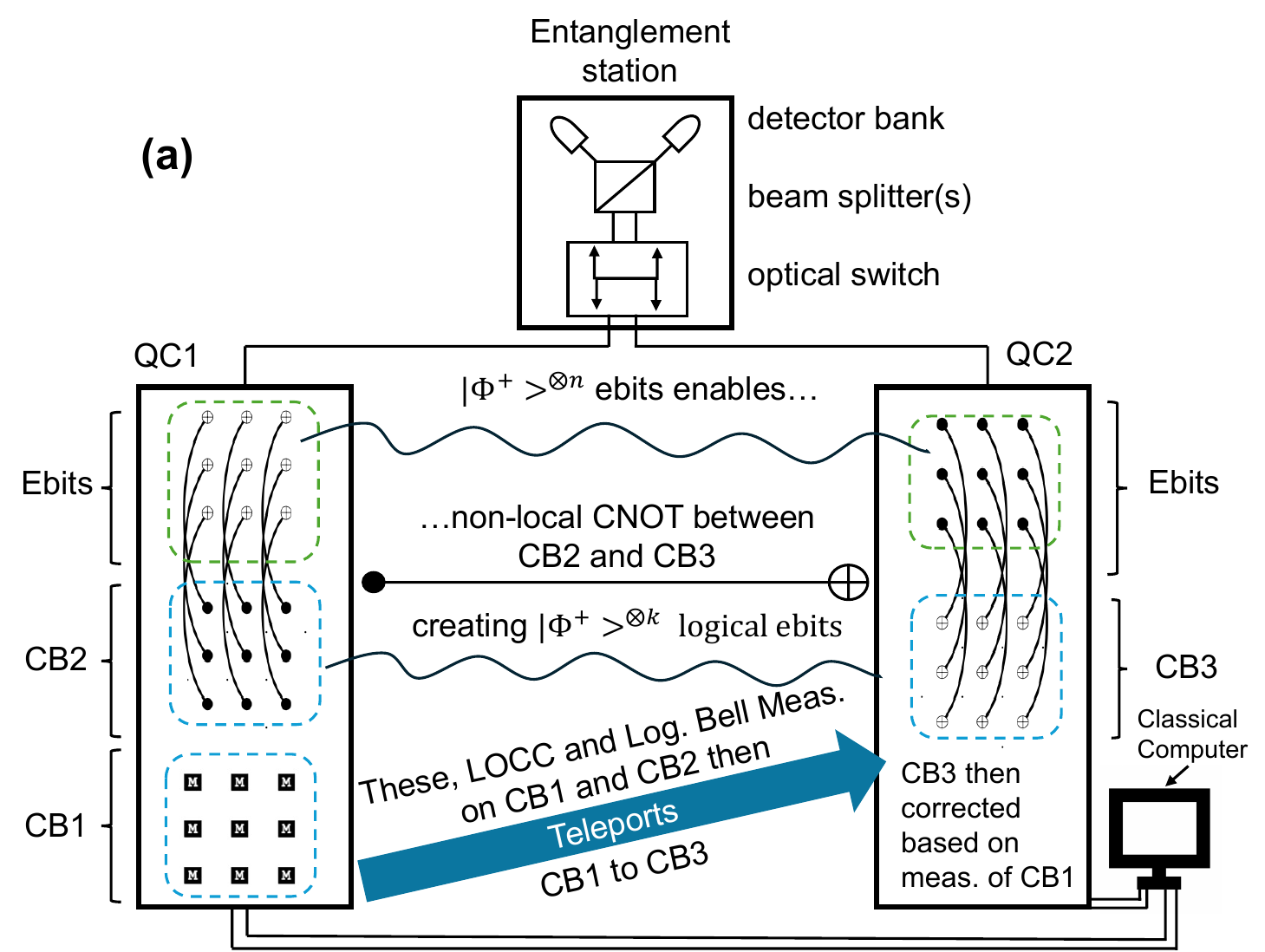}
    \label{fig:arch:a}
  \end{subfigure}\hfill
  \begin{subfigure}[t]{0.49\textwidth}
    \centering
    \includegraphics[width=\linewidth]{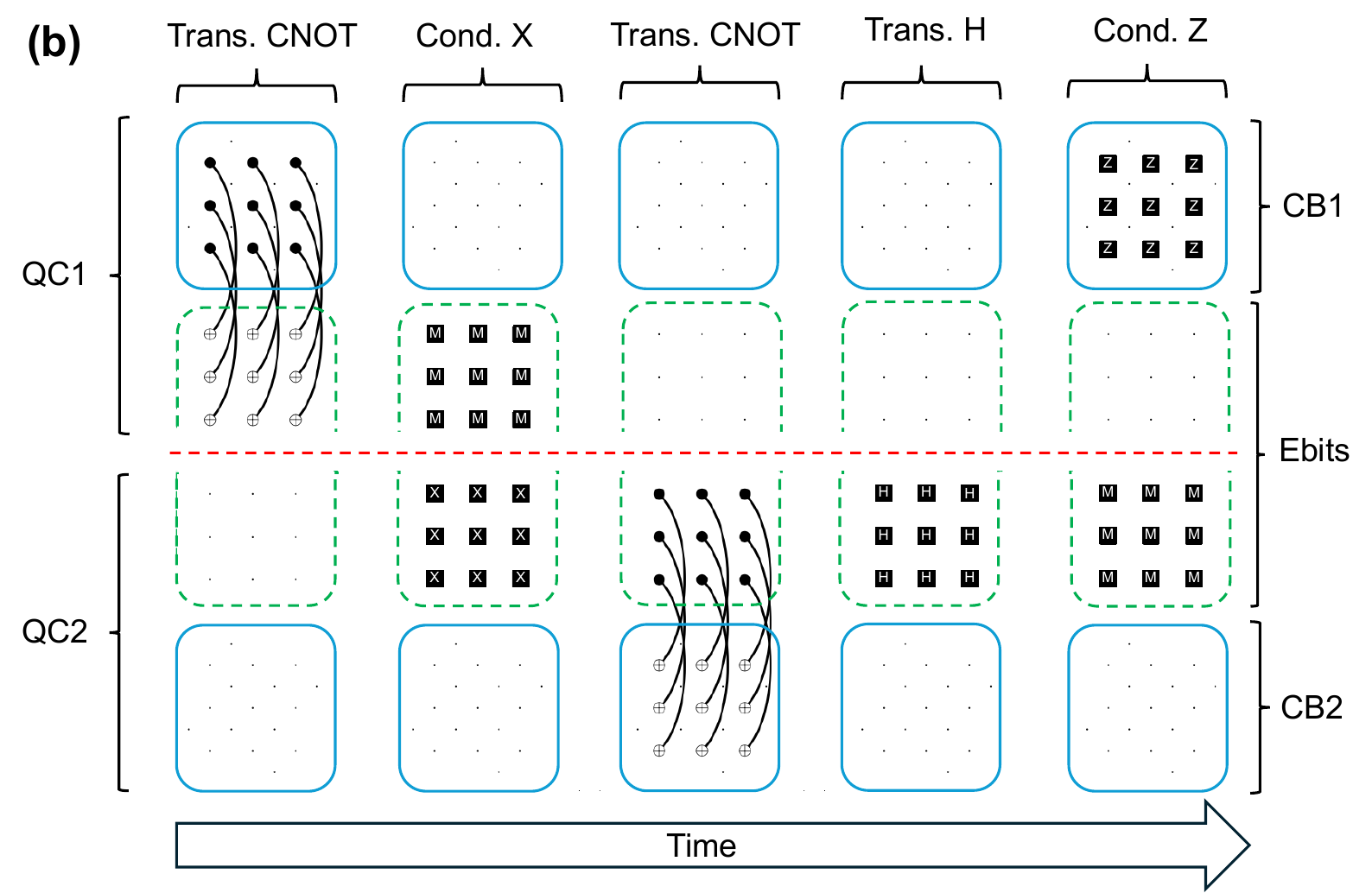}
    \label{fig:arch:b}
  \end{subfigure}
  \begin{subfigure}[t]{0.99\linewidth}
    \centering
      \includegraphics[width=0.95\linewidth]{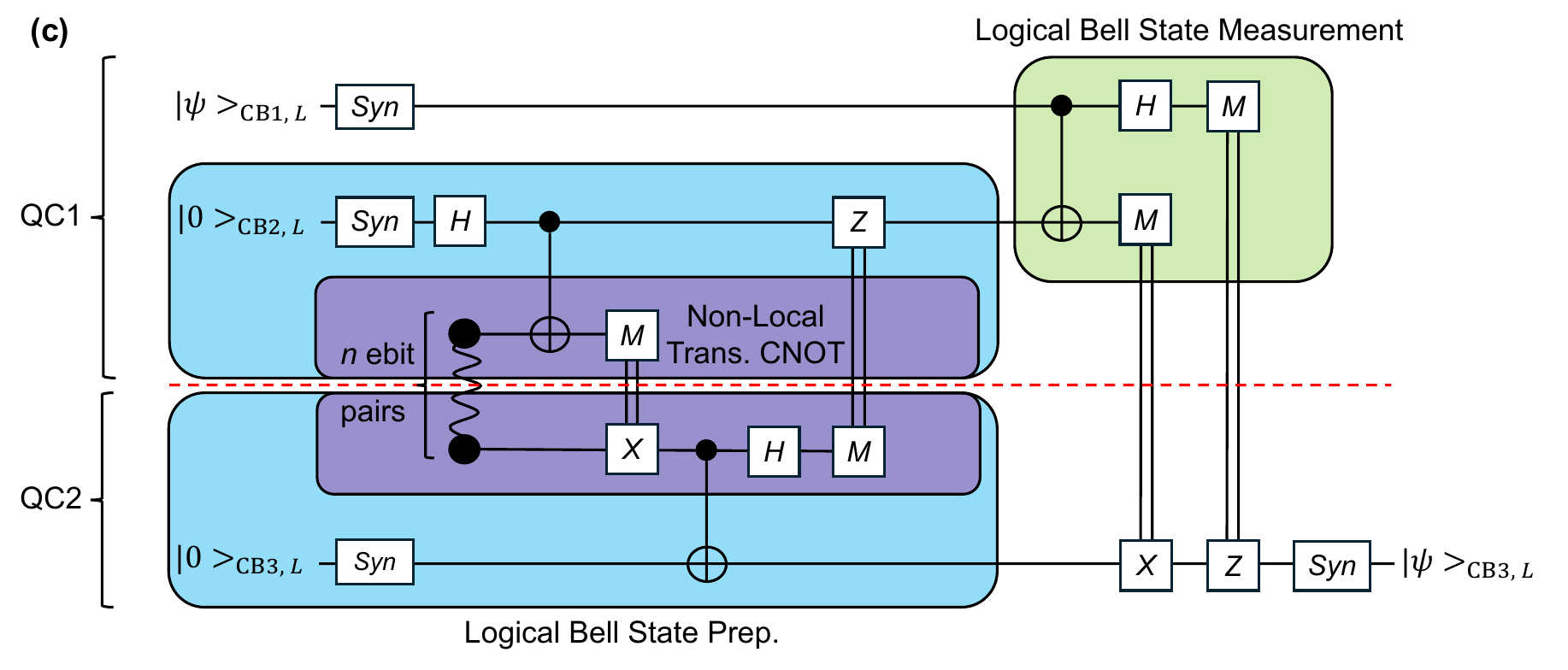}
    \label{fig:arch:c}
  \end{subfigure}
  \caption{\textbf{Architectural overview and implementation of
      distributed transversal operations.} \textbf{(a)} Nodes are
    interconnected via an entanglement station that contains an
    optical switch, beam splitter(s) and detectors, which enable the
    generation of ebits between nodes. This diagram illustrates how
    CB1 in QC1 can be teleported to CB3 in QC2. \textbf{(b)} Non-local
    transversal CNOT carried out between CB1 and CB2 located on QC1
    and QC2, respectively. The red line indicates the physical
    separation between devices. This operation involves a sequence of
    transversal CNOTs between code blocks and ebits, measurements and
    (conditional) single-qubit gates. Blue squares indicate logical
    code blocks. Green dashed lines indicate blocks of
    ebits. \textbf{(c)} A circuit for fault tolerant teleportation of
    logical code block $\ket{\psi}$ from quantum computer QC1 to QC2
    using two $\ket{0}$ code blocks, $n$ ebit pairs and local
    operations, only. The blue Logical Bell State Preparation
    subcircuit uses ebits to create $k$ logical Bell states between
    code block CB2 and CB3. This resource is then used to teleport CB1
    from QC1 to CB3 on QC2. There are syndrome extraction rounds
    (\emph{syn}) placed between certain operations, details of the
    exact numbers of syndrome extraction rounds are in the Methods
    section. }
  \label{fig:arch}
\end{figure*}

Quantum processors today operate with physical error rates that remain
too high for running deep quantum circuits without fault
tolerance. Quantum error correction (QEC) addresses this by encoding
logical qubits into 100s-1000s of physical qubits and repeatedly
extracting syndromes to suppress errors below the physical
rate. Achieving useful, fault-tolerant (FT) computation therefore
demands large-scale devices with thousands of qubits or more. At such
scales, single-chip systems face practical limits: fabrication yield,
wiring and connectivity, cooling power, optical addressing, and
constraints on trapped atoms or ions~\cite{constraints1}. These
considerations motivate distributed quantum computing (DQC):
interconnecting modules to act as a larger processor.

Several routes toward DQC exist. One line of work replaces quantum
links with classical post-processing, ``circuit knitting,'' which
partitions a circuit and recombines measurement data from separately
executed fragments~\cite{circuitKnitting}. Although appealing for
near-term devices, sampling overheads grow exponentially with the
number of cuts, restricting applicability to circuits with very
limited entanglement. Alternatively, one can couple modules
quantum-mechanically. Approaches include ion/atom shuttling between
chiplets~\cite{sussex1,prlTransport} and the creation of
inter-module entanglement resources (``ebits,'' i.e., Bell pairs
shared across nodes) used with local operations and classical
communication (LOCC). We focus on the ebit-based architecture because
it preserves modularity, works across heterogeneous hardware, and
supports natural FT primitives, non-local gates via gate teleportation
and logical state teleportation, without additional quantum
communication once ebits are
available~\cite{eisert_optimal_2000,tele}. Compilation choices then
determine where to place logical data and when to invoke non-local
operations~\cite{dqcCompile}.

Present-day ebits are substantially noisier than local two-qubit
gates: reported fidelities near 97\% imply error rates up to two
orders of magnitude worse than on-chip
operations~\cite{oxfordNAture,angLiArquin}. Entanglement distillation
can raise ebit fidelity at the cost of significant resource
overheads~\cite{harvard}. Understanding when imperfect ebits suffice,
and what logical distances are required, therefore becomes essential
for setting hardware targets and deciding if/when to distill.

Prior studies have largely examined distributed syndrome extraction or
lattice-surgery primitives rather than full logical operations across
nodes. Works using small GHZ-mediated nodes~\cite{singh,Nickerson}
show limited suppression under realistic ebit noise. Patch-extended or
seam-based surface-code layouts tolerate non-local gate noise up to an
order of magnitude higher than local
noise~\cite{Ben2,rametteNPJ,UCR}. Floquet-style distributed variants
also exhibit promising behavior in quantum memory
experiments~\cite{evanUCL}. Other experiments using small to medium
codes have demonstrated logical teleportation within a single device
using transversal gates~\cite{quantinuum, transversalTele} and/or
lattice surgery~\cite{quantinuum}. However, end-to-end, circuit-level
simulations of transversal non-local CNOT and full-block teleportation
between error-corrected nodes, especially with modern quantum
low-density parity check (qLDPC) codes, have remained largely
unexplored.

We present full-circuit simulations of
two transversal FT primitives executed between nodes of a DQC using
imperfect ebits: (i) a transversal non-local CNOT (gate teleportation)
between remote logical blocks, and (ii) transversal teleportation of
an entire logical code block. We benchmark two code families with
complementary trade-offs: The surface code~\cite{surfacecode} and
bivariate-bicycle (BB) qLDPC
codes~\cite{bravyi,wang2024coprimebivariatebicyclecodes,effNeutral}. Our
simulations operate in the low-noise regime, physical error rates down
to $10^{-5}$, to probe for potential error floors and to support
extrapolation to target logical error rates relevant for large-scale
algorithms such as Shor's algorithm (e.g., $p_L \sim
10^{-12}$)~\cite{gidney2025factor2048bitrsa}.

In addition, to evaluate the physical feasibility of generating the
required number of ebits in time to avoid their decoherence, we
perform circuit-level simulations of the effect of ebit decoherence on
non-local CNOT fidelity. We compare two ebit generation schemes:
one-at-a-time and $O(d)$ at a time, with the latter inspired by recent
experimental developments~\cite{UIUCPara}. This enables us to confirm
what ebit generation speeds are required given different device
decoherence rates.

Further, we compare distributed transversal CNOT operations to
distributed lattice surgery for executing Pauli string rotations. In
general, transversal approaches require all-to-all connectivity
whereas for lattice surgery operations nearest-neighbor connectivity
is sufficient. These are core operations required to execute a wide
variety of routines, ranging from Hamiltonian quantum simulation for
materials simulation~\cite{paulStringRef} to quantum phase
estimation~\cite{nielsen}, quantum linear systems
algorithms~\cite{HHL}, variational algorithms~\cite{vqeEtc} and
quantum signal processing~\cite{qsp}. We see how the parallelism
induced by qLDPC codes can minimize the ebits required per operation.

In performing these simulations, we developed code that enables
arbitrary transversal monolithic and/or distributed logical operations
between arbitrary numbers of either surface or BB code blocks. We name
this library Transversal Multiple Code Block Simulator (TMCBS). Our
library uniquely constructs full circuit-level models, generating
complete detector error models and parity-check matrices for entire
operations. 

By contrast, a coarse approximation of non-local operations is to
elevate or adjust the noise of cross-module
operations~\cite{routingCard, rametteNPJ}. This does not capture the
full complexity introduced by physical ebits, including their distinct
noise channels, the ancillary operations required for their use, and
the consequent propagation of errors across multiple code blocks,
which significantly increases decoding complexity (see direct link
vs. gate teleportation in~\cite{UCR} for an analysis of this in a
lattice surgery based context).

Our library captures this complexity
and enables more complex experimentation in terms of ebit noise. The
TMCBS library will be publicly available after publication and
supports MPI parallelism, enabling runs on hundreds to $>10^3$ CPU
cores across multiple nodes on compute clusters. This enabled us to
decode large codeblocks and logical circuits.

Our work contributes actionable design rules for distributed
fault-tolerant computation. First, we establish that transversal
non-local CNOTs should be preferred over logical teleportation, with
the latter reserved for cases where it removes on the order of
10 non-local CNOTs at the circuit level. Second, we translate
these findings into concrete build targets: At a physical error rate
of $\sim 10^{-3}$ with inter-node ebit noise of $\sim 10^{-2}$, a
surface-code distance of $d \approx 29$ suffices to achieve
per-operation logical error rates near $10^{-12}$ for the non-local
CNOT, enabling large-scale distributed algorithms. At a physical error rate of $\sim 10^{-4}$ with inter-node ebit noise of $\sim 10^{-3}$, a
surface-code distance of only $d \approx 11$ suffices to achieve
per-operation logical error rates near $10^{-12}$. Third, we find that
parallelism in ebit generation relaxes the ebit generation speed
required for minimal decoherence noise by an order of magnitude,
entering the experimentally feasible regime for certain
platforms. Lastly, benchmarking against distributed lattice surgery
shows that qLDPC based transversal approaches on the architectures
that can efficiently support them can have significantly lower
effective LER, fewer ebits required per logical operation and faster
execution. Collectively, these contributions link architecture,
error-correction parameters, and compiler policy into a coherent
blueprint for scalable distributed quantum computing.

\section{Results}
\label{sec:results}

\subsection{Architecture and operations}

In initially evaluating the two primitives, we assume a two-node
architecture, where ebits can be created between the nodes (see
Fig.~\ref{fig:arch}). This is a general model agnostic of
architectural details (connectivity, movement cost etc.). This is to
keep the results of the analysis widely applicable and ensure the
focus is on analyzing the effect of ebit noise for FT non-local
operations and teleportation realized from purely transversal
operations. Later in this work we consider more than two nodes and the
impact of decoherence on ebits due to their production time.

In general, transversal operations are implemented by carrying out the
logical operation on each physical qubit: A transversal $X$ gate is
applied to a codeblock by applying a physical $X$ gate to every
physical qubit, a transversal CNOT is applied to a pair of codeblocks
by executing a physical CNOT between the corresponding pairs of
physical qubits across the two codeblocks.

Transversal operations can be implemented quickly and efficiently on
certain classes of quantum computers, e.g., on a neutral atom device,
where acousto-optic deflectors (AODs) can control a large number of
physical qubits simultaneously~\cite{lukinNature}. That being said,
the requirement that every qubit/ebit in a control codeblock must
interact with their corresponding qubits/ebits in the target codeblock
makes transversal operations significantly more difficult to
efficiently implement on some platforms, e.g., on nearest-neighbor
limited superconducting or silicon platforms. Our conclusions
therefore apply to platforms that can efficiently provide this
capability such as neutral atom devices~\cite{lukinNature} and
potentially trapped ion devices~\cite{quantinuum}.

In this work, we implement fault-tolerant CNOTs and state
teleportation using only transversal gates, measurements, and ebits.
For either operation on $\llbracket n,k,d\rrbracket$ codeblocks, $n$
ebit pairs are used. Fig.~\ref{fig:arch} shows how the operations can
be implemented between codeblocks located on different devices.

\subsection{Circuit-level distance of distributed primitives}
\begin{table}[t]
  \caption{\label{tab:bound}%
    \textbf{Upper bound on circuit-level distance for different SC and
      BB code blocks.} The circuits in question are the non-local CNOT
    circuit involving 2 codeblocks and the full teleportation circuit
    involving 3 codeblocks. A dash indicates a result could not be
    found in feasible time or the heuristic reported a circuit
    distance higher than code distance.}
\begin{ruledtabular}
\begin{tabular}{lcr}
\textrm{\textbf{Code}}&
\textrm{$\textbf{\emph{d}}$ \textbf{non-local}}&
\textrm{$\textbf{\emph{d}}$ \textbf{teleport}}\\
\colrule
$\llbracket 18,4,4\rrbracket$ BB~\cite{wang2024coprimebivariatebicyclecodes}& $3$ & $3$ \\
$\llbracket 36,4,6\rrbracket$ BB~\cite{wang2024coprimebivariatebicyclecodes}& $6$ & $6$ \\
$\llbracket 54,4,8\rrbracket$ BB~\cite{wang2024coprimebivariatebicyclecodes}& $7$ & $8$ \\

$\llbracket 90,8,10\rrbracket$ BB~\cite{bravyi}& $9$ & $10$ \\

$\llbracket 144,12,12\rrbracket$ BB~\cite{bravyi}& - & - \\

$\llbracket 9,1,3\rrbracket$ SC & $3$ & $3$ \\

$\llbracket 25,1,5\rrbracket$ SC & $5$ & $5$ \\

$\llbracket 49,1,7\rrbracket$ SC & $7$ & $7$ \\

$\llbracket 81,1,9\rrbracket$ SC & $9$ & $9$\\

$\llbracket 121,1,11\rrbracket$ SC & $11$ & $11$
\end{tabular}
\end{ruledtabular}
\end{table}

\begin{figure*}[t]
\centering \vspace{0pt} 
  \centering
  \includegraphics[width=0.95\linewidth]{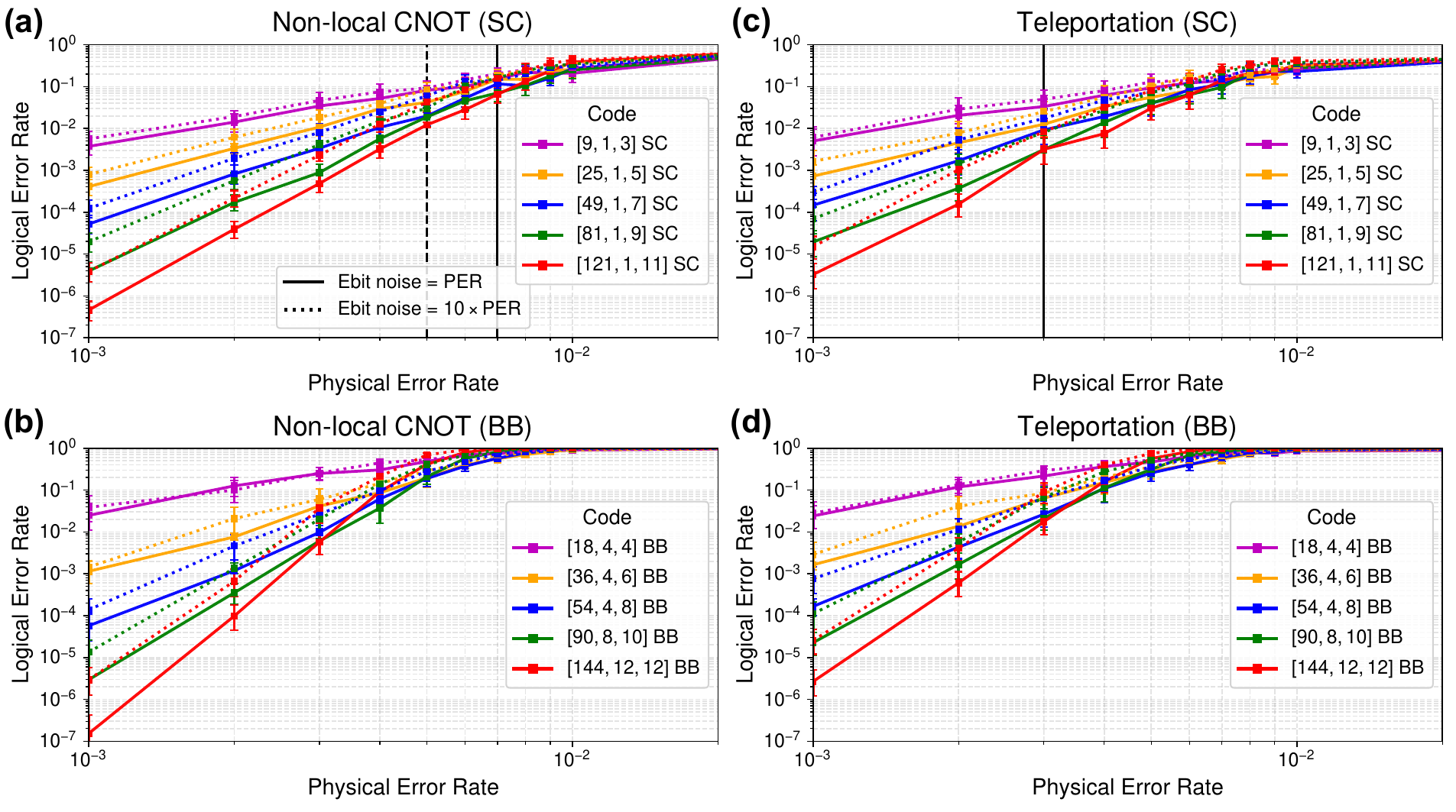}
  \caption{\textbf{Threshold plots for distributed transversal
      operations.} PER against LER for fixed ebit noise equal to PER
    (solid) or 10 times worse than PER (dashed). Thresholds are
    indicated by the vertical lines. \textbf{(a)} Non-local CX using
    surface codes $[d^2,1,d]$. \textbf{(b)} Non-local CX using
    selected BB codes that have identical or similar circuit level
    distance to the surface codes we consider. \textbf{(c)}
    Teleportation using SC codes. \textbf{(d)} Teleportation using BB
    codes.}
  \label{fig:thresh}
\end{figure*}

We first assessed the intrinsic fault-tolerance of our two distributed
primitives, a non-local, transversal CNOT between codeblocks located
on different devices and a full logical teleportation between devices,
by estimating their circuit-level distance via the \texttt{STIM}
undetectable-error search heuristic
(\texttt{search\_for\_undetectable\_logical\_errors}). Code distance
is the minimum weight of any Pauli operator that implements a logical
operation (in error or desired, see Supplementary
Note 1). Circuit distance is the minimum number of errors in a
logical circuit across space and time that causes an undesired logical operation on an observable (typically a logical qubit). In both cases, larger distances
lead to lower LERs. Table~\ref{tab:bound} reports upper bounds for
several surface-code (SC) and bivariate-bicycle (BB) instantiations
while holding the circuit templates fixed.

It is important to note that code distance does not necessarily
directly translate to circuit distance, especially for codes more
complex than the surface code~\cite{bravyi}. This is because the
syndrome extraction circuit used to infer errors and correct them via
decoding (Methods) can itself introduce errors into the code, e.g.,
via Hook errors~\cite{hookErrors} where errors on the ancilla qubits
enter the data qubits via the syndrome extraction circuit. Whilst the
syndrome extraction circuit for surface codes is circuit-distance
preserving, for more complex codes it is often difficult to design a
circuit that is both efficient and perfectly distance
preserving~\cite{bravyi}.

It is even less obvious that logical circuits like those considered in
this work are code-distance preserving. This is because how errors
propagate across and within multiple codeblocks is
non-trivial~\cite{duke, ghost}.

We find that for SC based circuits, the circuit-level distance matches
the code distance $d$ for both primitives, consistent with the SC's
distance-preserving syndrome-extraction circuit. In contrast, most of the BB codes we study 
have circuit-level distances below $d$. Both BB primitives are
affected by the fact the syndrome extraction circuits available in the
literature, including the one we use, are not perfectly
distance-preserving~\cite{bravyi}. 

Regardless, both primitives remain fault tolerant across all codes
probed: Increasing the code distance monotonically increases the
circuit-level distance upper bound for each primitive. Taken together,
Table~\ref{tab:bound} provides an end-to-end confirmation that fully
distributed logical operations (not just memory experiments) can
inherit distance from the underlying code family when implemented
transversally.

\subsection{Distributed operations thresholds and scaling}
\label{subsec:thresholds}

\begin{table}[t]
    \centering
     \caption{\textbf{Distances required to achieve certain logical error
      rates for different distributed logical operations.} Non-local
      CNOT circuit (NL) and the teleportation circuit (T). Ebit noise
      ratios indicated by $r$. Distance ($d$) is calculated using
      Eq.~\ref{eq:scaling} and Eq.~\ref{eq:distance_formula} and rounded
      to the nearest odd number.} 
    \label{tab:thresh}
    \begin{ruledtabular}
    \begin{tabular}{lccccc}
\textrm{} &\textrm{} &\textrm{} & \multicolumn{3}{c}{\textrm{$\bm{d}$ \textbf{for Target LER}}} \\
\textbf{Circuit} & \textbf{PER} & $\bm{r}$ & $\bm{10^{-8}}$ & $\bm{10^{-10}}$ & $\bm{10^{-12}}$ \\
      \colrule
NL & $10^{-3}$ &   1    & 15 & 19 & 25 \\
NL & $10^{-3}$ &   10    & 19 & 25 & 29 \\
NL & $10^{-4}$ &   1    & 7 & 9 & 11 \\
NL & $10^{-4}$ &   10    & 7 & 9 & 11 \\
T  & $10^{-3}$& 1& 21 & 29 & 39 \\
T  & $10^{-3}$& 10& 25 & 33 & 41 \\
T  & $10^{-4}$& 1& 7 & 9 & 13\\
T  & $10^{-4}$& 10& 7 & 11 & 13 \\
\end{tabular}
\end{ruledtabular}
\end{table}

\begin{table*}[t]
\caption{
\textbf{Overview of
  gate counts and parity check matrix sizes for different distributed
  operations and codes.}  One (1q) and two qubit (2q) gates and
  measurements (M) used by a logical non-local CNOT and teleportation
  circuits involving two and three instances of codeblocks
  respectively for different codes. The size of the space-time PCM
  generated by the STIM DEM is also shown, where each row corresponds
  to a detector in the circuit and each column corresponds to a
  distinct error mechanism within the circuit. The number of rows and
  columns (cols) is shown in the table.
}
\label{tab:cnotAndTele}
\begin{ruledtabular}
\begin{tabular}{l rrrrr @{\hspace{1.2em}} rrrrr}

\textrm{} & \multicolumn{5}{c}{\textrm{\textbf{Non-local CNOT}}} &
\multicolumn{5}{c}{\textrm{\textbf{Teleportation}}} \\

\textrm{\textbf{Code}} & \textrm{\textbf{1q}} & \textrm{\textbf{2q}} & \textrm{\textbf{M.}} & \textrm{\textbf{rows}} & \textrm{\textbf{cols}}
& \textrm{\textbf{1q}} & \textrm{\textbf{2q}} & \textrm{\textbf{M.}} & \textrm{\textbf{rows}} & \textrm{\textbf{cols}} \\
\colrule
$\llbracket 18,4,4\rrbracket$ BB        &   54 &  1548 &  288 &  144 &  1314 &  126 &  1674 &  342 &  144 &  1314 \\

$\llbracket 36,4,6\rrbracket$ BB        &   108 &  3096 &  576 &  288 &  2628 &  252 &  3348 &  684 &  288 &  2646 \\

$\llbracket 54,4,8\rrbracket$ BB        &  162 &  4644 &  864 &  432 &  3942 &  378 &  5022 & 1026 &  432 &  3969 \\

$\llbracket 90,8,10\rrbracket$ BB        &   270 &  7740 &  1440 &  720 &  6570 &  630 &  8370 &  1710 &  720 &  6615 \\

$\llbracket 144,12,12\rrbracket$ BB     &  432 & 12,384 & 2304 & 1152 & 10,512 & 1008 & 13,392 & 2736 & 1152 & 10,584 \\

$\llbracket 9,1,3\rrbracket$ SC        &   27 &  354 &  130 &  64 &   245 &  63 &  387 &  156 &  64 &  238 \\

$\llbracket 25,1,5\rrbracket$ SC        &   75 &  1170 &  386 &  192 &   847 &  175 &  1275 &  460 &  192 &  832 \\

$\llbracket 49,1,7\rrbracket$ SC        &  147 &  2450 &  770 &  384 &  1809 &  343 &  2667 & 916 &  384 &  1782 \\

$\llbracket 81,1,9\rrbracket$ SC        &  243 &  4194 &  1282 &  640 &  3131 &  567 &  4563 & 1524 &  640 &  3088 \\

$\llbracket 121,1,11\rrbracket$ SC      &  363 &  6402 & 1922 &  960 &  4813 &  847 &  6963 & 2284 & 960 &  4750 \\
\end{tabular}
\end{ruledtabular}
\end{table*}

\begin{figure*}[!h]
  \includegraphics[width=0.87 \linewidth]{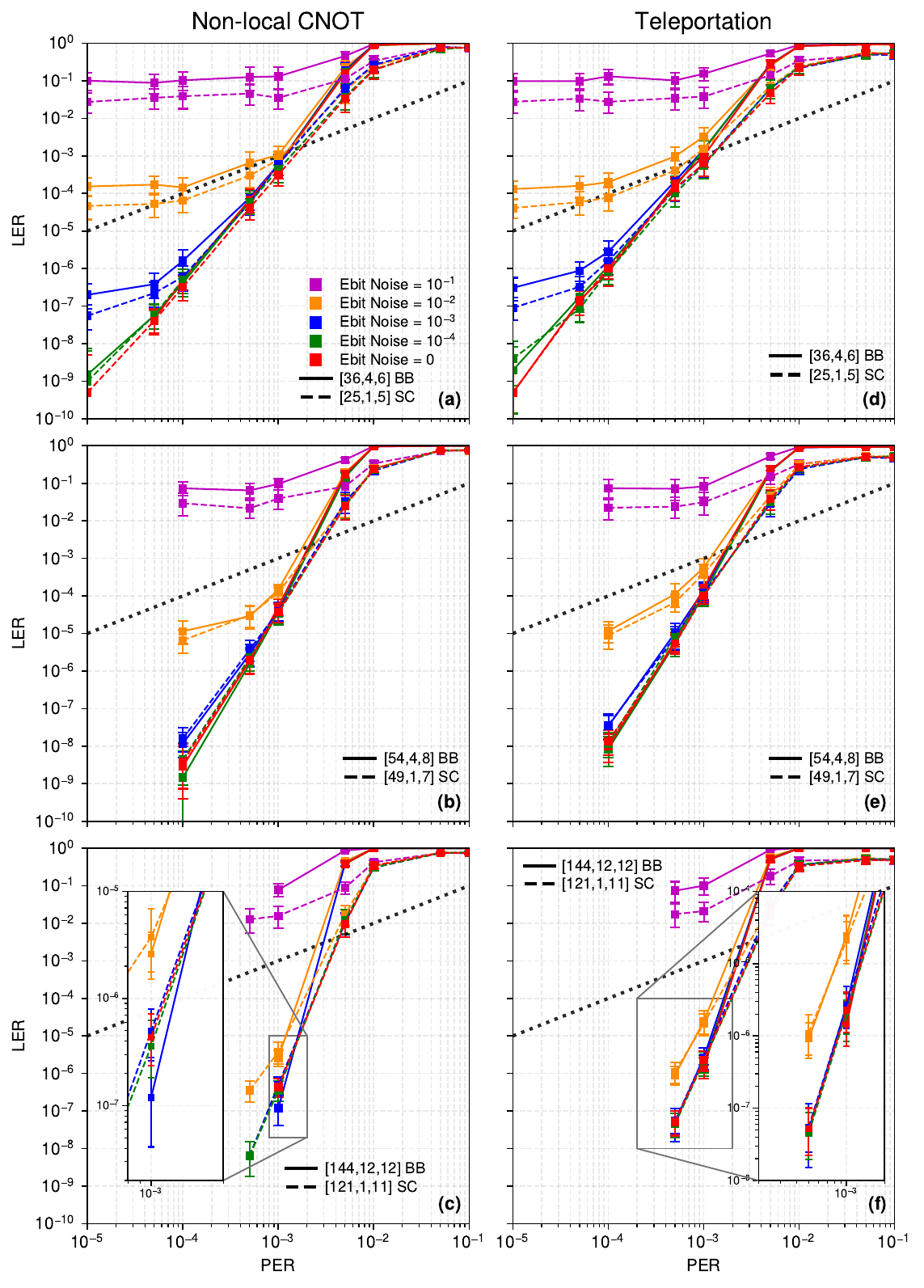}
  \caption{\textbf{Performance of distributed operations for different
    ebit error rates.} Physical noise rate against LER for different
    ebit noise rates for non-local CNOT \textbf{(a-c)} or teleportation \textbf{(d-f)}
    between pairs of SC or BB codeblocks. Solid lines indicate BB
    codes, dashed lines indicate SC codes.}
  \label{fig:all}
\end{figure*}

Next, we extracted operational thresholds for each primitive under
circuit-level depolarizing noise on one- and two-qubit gates,
state-preparation and measurement (SPAM), and on heralded ebit
generation. Given that we are evaluating a circuit rather
than a memory experiment, we provide a loose definition of a threshold
as the physical error rate (PER), where a monotonic increase in
distance results in a monotonic decrease in LER. A higher threshold is
therefore better than a lower threshold. We attempt to calculate
thresholds only for the surface code due to the heterogeneity in the
BB codes we use in the rest of this paper. Fig.~\ref{fig:thresh}(a-c)
show the surface-code thresholds for two regimes: (i) ebit error rate
equal to the physical error rate ($p_{\text{ebit}}=p$) and (ii)
$p_{\text{ebit}}=10p$.

We observe a clear separation between the two primitives. For the
non-local CNOT, the operational threshold lies at
$p_{\mathrm{th}}\!\approx\!7\times10^{-3}$ when $p_{\text{ebit}}=p$
and at $p_{\mathrm{th}}\!\approx\!5\times10^{-3}$ for
$p_{\text{ebit}}=10p$. For the full teleportation, the corresponding
thresholds are lower and coincide at $\approx 3\times 10^{-3}$ for
both ebit-noise settings.

This gap is consistent with the additional codeblocks,
gates, basis changes, measurements and conditional operations in teleportation (Fig.1(c)), which increase the
number of errors and error mechanisms, but more importantly significantly increase the complexity of the de-coding task~\cite{duke, ghost}.

To project finite-distance data to target LERs, we apply the usual
deep-subthreshold scaling form (Methods) to our surface-code data
yielding the extrapolations in Table~\ref{tab:thresh}. For a non-local
CNOT at $p=10^{-3}$ with $p_{\text{ebit}}=10^{-2}$, a LER of
$10^{-12}$ is achieved at distance $d\approx29$. For teleportation at
same $p$ and $p_{\text{ebit}}$, $d\approx41$
is necessary for the same $10^{-12}$ LER. This LER is cited as sufficient
to execute large-scale algorithms such as Shor's
algorithm~\cite{gidney2025factor2048bitrsa}. At $p=10^{-4}$, the distances required are significantly lower. 

Whilst we do not attempt to identify a threshold, we show a similar
case in Fig.2(b) for a set of BB codes that have the same
circuit-level distance as the surface codes considered. We see that
the LER for a given PER and distance is very similar. Given this,
Table~\ref{tab:thresh} can potentially be used as a rough guide for BB
codes too.

This similarity in performance was not the case under BP-OSD
(Supplementary Note 7). The decoder we use,
Tesseract~\cite{teser}, has been shown in~\cite{teser} to be better
than BP-OSD at decoding BB codes. But the almost identical performance
between same circuit distance SC and BB codes on logical, especially
distributed, has not been previously seen.

This is especially interesting given the significantly increased
complexity of BB codes and their syndrome extraction circuits (see
Tab.~\ref{tab:cnotAndTele}). This shows that for distributed quantum
computing qLDPC codes can enable a roughly similar number of ebits to
be used to carry out multiple logical operations at a similar PER
compared to just one logical operation using surface codes. We discuss
this further later in the text.

\subsection{Non-local CNOT performance}

Figures~\ref{fig:all}(a-c) compare the distributed CNOT across small,
medium, and larger codes for multiple ebit-noise ratios. The dotted
diagonal marks the "break-even" line where LER equals PER; points
beneath it indicate a net benefit from error correction. For small
codes of comparable $n$, both SC and BB reach a break-even point
around $p\approx 10^{-3}$ provided $p_{\text{ebit}}\lesssim
10^{-3}$. In general, as we increase the distance from small to medium
to large, corresponding and expected increases in the steepness of
the slopes are observed. At high PERs ($\geq10^{-3}$) BB codes have
worse LER than SC. But below this their performance seems equivalent
to SC. These graphs show that the logical non-local CNOT works
appropriately: Lower PERs and higher code distances enable the
operations to be carried out with lower LERs with no obvious error
floors and a clear path to very low LERs. In addition, we observe that
ebits up to 10 times worse than PER can be tolerated. Ebits worse than
this level cause flat-lining in LER.

It should be noted that multiple non-local CNOTs that use the same
control qubit can be carried out without the need to generate more
ebits. After the measurement of one half of the ebit pairs, the
remaining half mirrors the state of the logical codeblock on the other
device. The remaining ebits, or original codeblock, can be used
indefinitely as control qubit(s) on their device until a non-diagonal
gate is applied to the codeblock or ebits~\cite{felix}.

Empirically, this property of non-local CNOTs means that in
distributed implementations of common circuits/gadgets the number of
codeblocks worth of ebits needed can be between 61\% and 85\% lower
than the number of non-local gates~\cite{felix}.

\subsection{Logical teleportation performance}

Figures~\ref{fig:all}(d-f) repeat the analysis for full logical
teleportation.  The qualitative ordering across code families, sizes,
and ebit-noise ratios is unchanged, but the absolute performance
shifts: At matched $(p,p_{\mathrm{ebit}})$ and distance, teleportation
yields a logical error rate up to an order of magnitude
higher than the corresponding transversal non-local CNOT.

The explanation for this is non-obvious. For the surface code, the circuit-level distance is identical for the two primitives and for BB codes it is similar. Whilst the number of one-qubit gates, two-qubit gates, and measurements are all higher for the teleportation primitive (in the case of the latter two, by $\leq10\%$), the dimensions of the PCM are nearly identical between the two primitives (see Tab. III). In addition, BB instances under non-local CNOT can have significantly larger two-qubit gate counts and PCMs than surface-code teleportation circuits at similar distance, yet have significantly lower LER. These comparisons indicate that simple proxies such as circuit distance, two-qubit gate count, and DEM width are not sufficient to explain the teleportation penalty.

This underscores the value of full circuit-level simulation for
distributed primitives: Protocols that appear comparable under such
metrics can nonetheless exhibit markedly different logical
performance.

We attribute the gap primarily to qualitative differences in circuit
structure, additional entangling steps and measurement-conditioned
feed-forward in teleportation, which introduce correlated fault
mechanisms across codeblocks, increasing the decoding burden in ways
that are not captured by aggregate metrics such as distance or gate
count.  Consistent with this interpretation, later results show that
decoding becomes substantially more challenging as logical circuits
span multiple codeblocks, suggesting that quantum memory benchmarks
alone are an incomplete proxy for decoder performance in distributed
settings.

Nevertheless, despite the larger surface for error propagation, the
LER-versus-$p$ slopes in Fig.~\ref{fig:all}(d-f) remain steep in the
subthreshold regime with LERs $\leq 10^{-7}$ reached at PER $5\times10^{-4}$.

\subsection{Delay noise}
\label{sec:delay_noise}

\begin{figure*}[t]
  \centering
    \includegraphics[width=1 \linewidth]{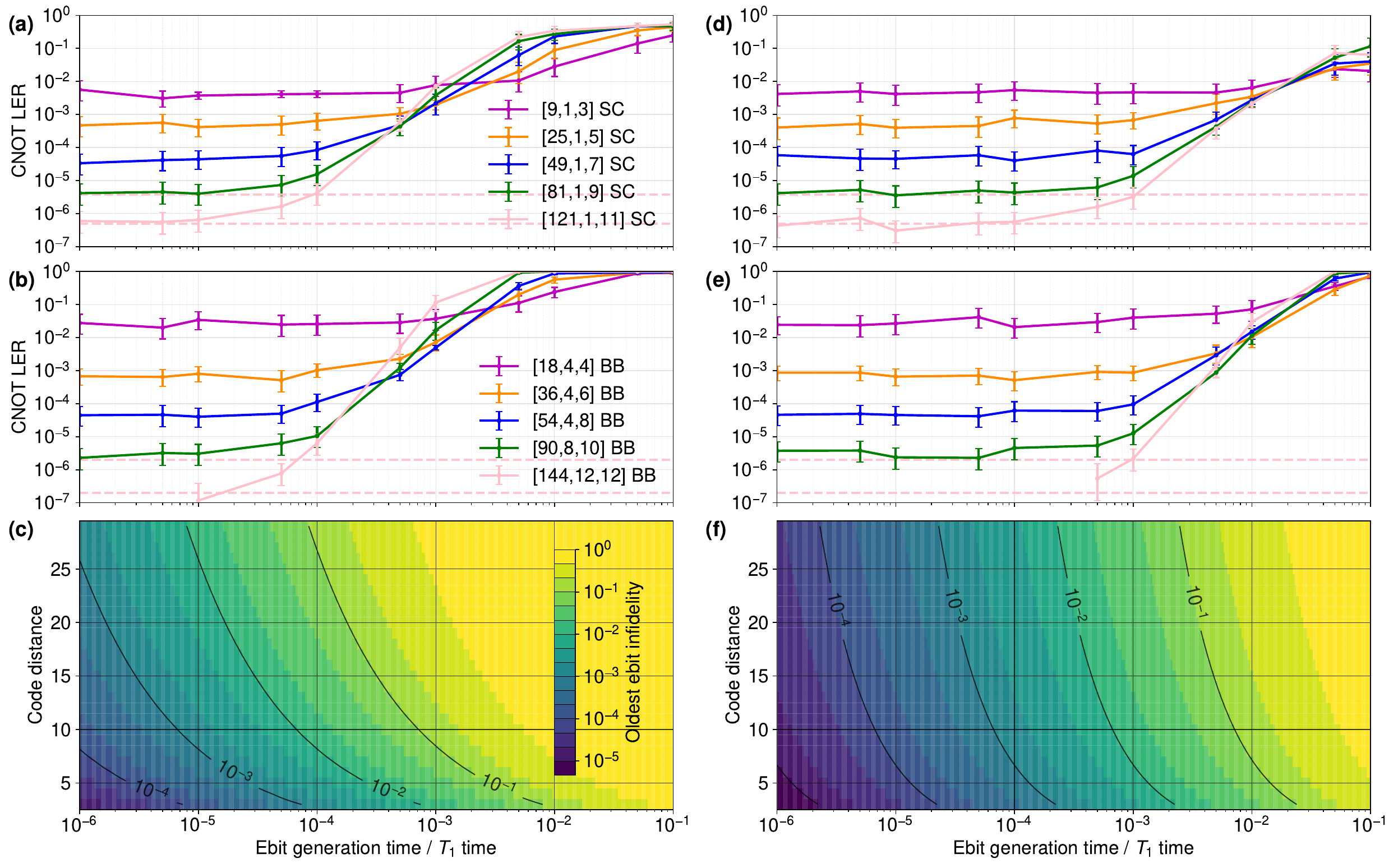}

  \caption{\textbf{Effect of ebit decoherence on logical error rate
  for non-local CNOT.} PER and ebit noise set to $10^{-3}$. Horizontal
  lines indicate decoherence free LER for $p_{ebit}=p$ and
  $p_{ebit}=10p$ for the largest codes. Heatmap colors show
  oldest-ebit infidelity, discretized into logarithmic 1–2–5 bins per
  decade. Contour lines mark constant infidelity with labels in powers
  of ten. \textbf{(a-c)} One-at-a-time generation method. \textbf{(a)} Surface codes, \textbf{(b)}
  BB codes, \textbf{(c)} Heatmap of the infidelity of the oldest ebit generated
  immediately before execution of the non-local CNOT for different code
  distances d for $n=d^2$. \textbf{(d-f)} Line generation method. \textbf{(d)} Surface
  codes, \textbf{(e)} BB codes, \textbf{(f)} Heatmap of the infidelity of the oldest ebit
  line immediately before execution of the non-local CNOT for
  different code distances $d$ for $n=d^2$.  }

  \label{fig:delayNoise}
\end{figure*}

Ebit generation latency can be non-negligible compared to the
coherence times of current hardware, particularly for transversal
distributed primitives, which consume $n$ ebits for
$\llbracket n,k,d\rrbracket$ codeblocks.  This means that the
decoherence accumulated by ebits that must wait in memory until the
full set of $n$ is available for transversal operations can become a
significant source of error in addition to ebit infidelity.  To
quantify when this waiting-time contribution becomes relevant and to
extract minimum requirements on the ebit generation rate for a given
hardware coherence budget, we perform circuit-level simulations in
which each ebit is subjected to decoherence noise determined by its
age prior to use.

We model each ebit as being generated at some time before it
participates in the logical circuit and apply a decoherence channel to
both halves of the ebit for an idle duration equal to the ebit's age
(Methods).

We consider two idealized ebit production schedules: (i) one-at-a-time
generation, where ebits are produced sequentially, and (ii)
parallel-line generation, where $\mathcal{O}(d)$ ebits are produced
simultaneously per time step (motivated by architectures capable of
parallel photonic links or multiple simultaneous entanglement
channels~\cite{UIUCPara}).

By combining the results of circuit level simulations in
Fig.~\ref{fig:delayNoise}(a,b,d,e) with an analytical sweeping of
decoherence noise on the oldest qubits
(Fig.~\ref{fig:delayNoise}(c,f)) we can identify a simple design
rule. We find that ebit wait time becomes a non-negligible factor when
the memory noise on the oldest ebits becomes comparable to the
baseline physical error rate. We do not observe a significant
difference between how BB and SC codes are affected by decoherence.

For the one-at-a-time generation regime, distance-25 codes require a
generation-rate to decoherence-rate ratio of $10^{-5}$ and $10^{-6}$
for $10^{-2}$ and $10^{-3}$ decoherence noise, respectively. Codes of
roughly $d=11$ need about an order of magnitude less than this. For
platforms with coherence times on the order of seconds, such as
neutral atoms~\cite{NAt1} and trapped ions~\cite{quantT1}, this
translates to required ebit generation rates on the order of
microseconds.  The best reported ebit generation rate is $4 \times
10^{-3}$ seconds per ebit in~\cite{monroe}, where a roadmap to kHz rates ($\leq
10^{-3}$ seconds per ebit) was reported. Whilst
challenging, longer coherence times on the order of 10s-100s on these
platforms are not intrinsically impossible~\cite{ NCTIT1}, so the
one-at-a-time approach is within feasible experimental
reach. Conversely, for platforms with microsecond decoherence times,
such as superconducting qubits, required ebit generation rates would
be on the order of nanoseconds for the largest codes, which is likely experimentally
infeasible.

With parallel generation schemes of $\mathcal{O}(d)$ ebits at a time,
the required speeds drop by an order of magnitude. This makes the
requirements for trapped ion and neutral atom platforms more feasible
for the largest codes ($10^{-4}$ to $10^{-5}$) and large codes
($10^{-3}$ to $10^{-4}$), but still infeasible for superconducting
platforms.

The ideal generation scheme would involve generating all
required ebits simultaneously, which would make distributed
transversal schemes highly efficient on atomic platforms and
potentially viable on superconducting platforms.

\begin{figure*}[t]
  \centering

  \begin{minipage}[t]{0.40\textwidth}
    \vspace{2.5cm} 
    \centering
    \includegraphics[width=\linewidth]{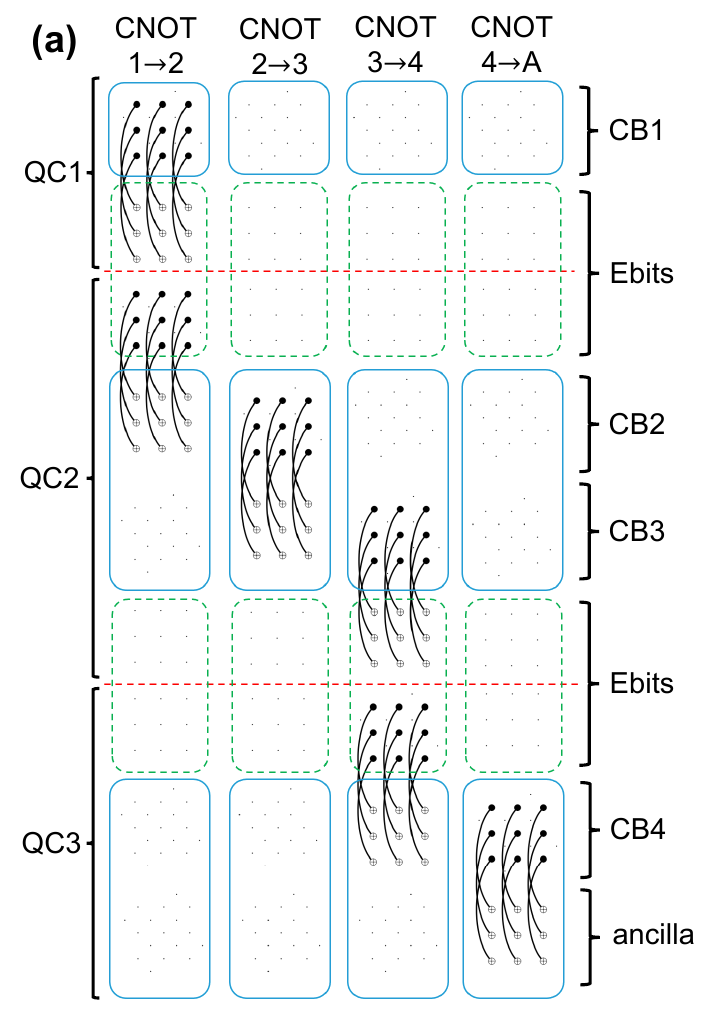}
  \end{minipage}\hfill
  \begin{minipage}[t]{0.60\textwidth}
    \vspace{1cm} 
    \centering
    \includegraphics[width=\linewidth]{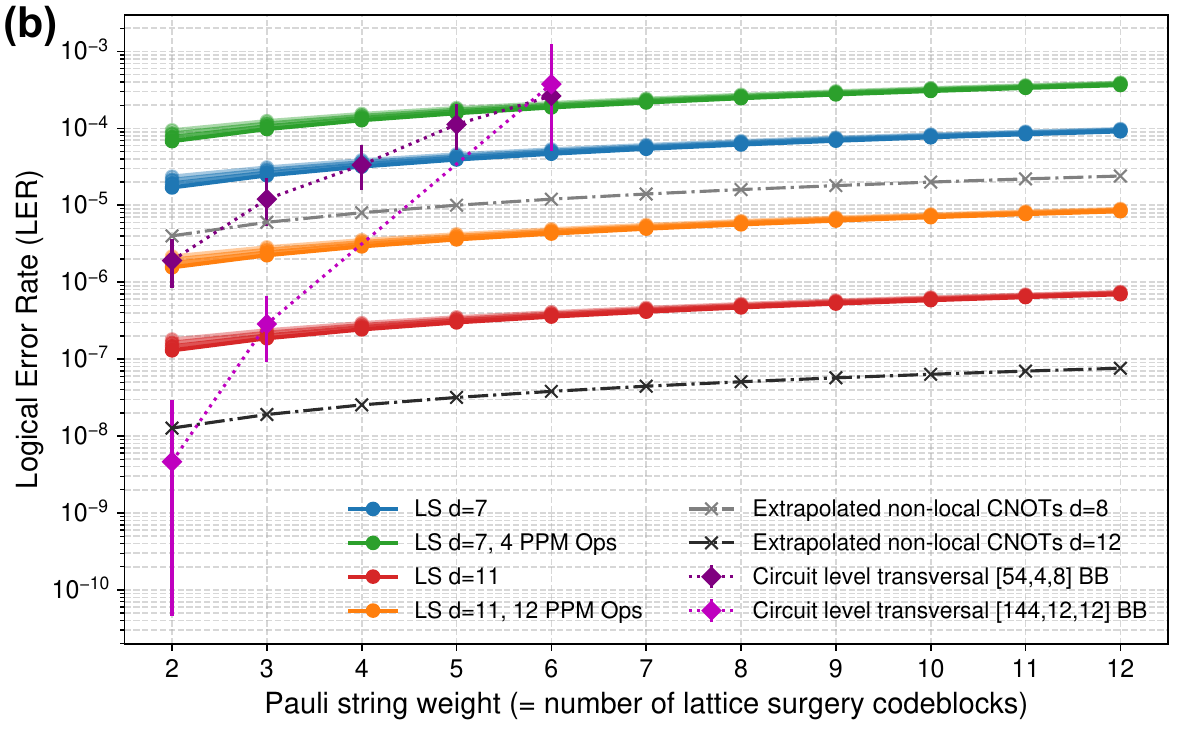}

    \vspace{0.8em} 

    \includegraphics[width=\linewidth]{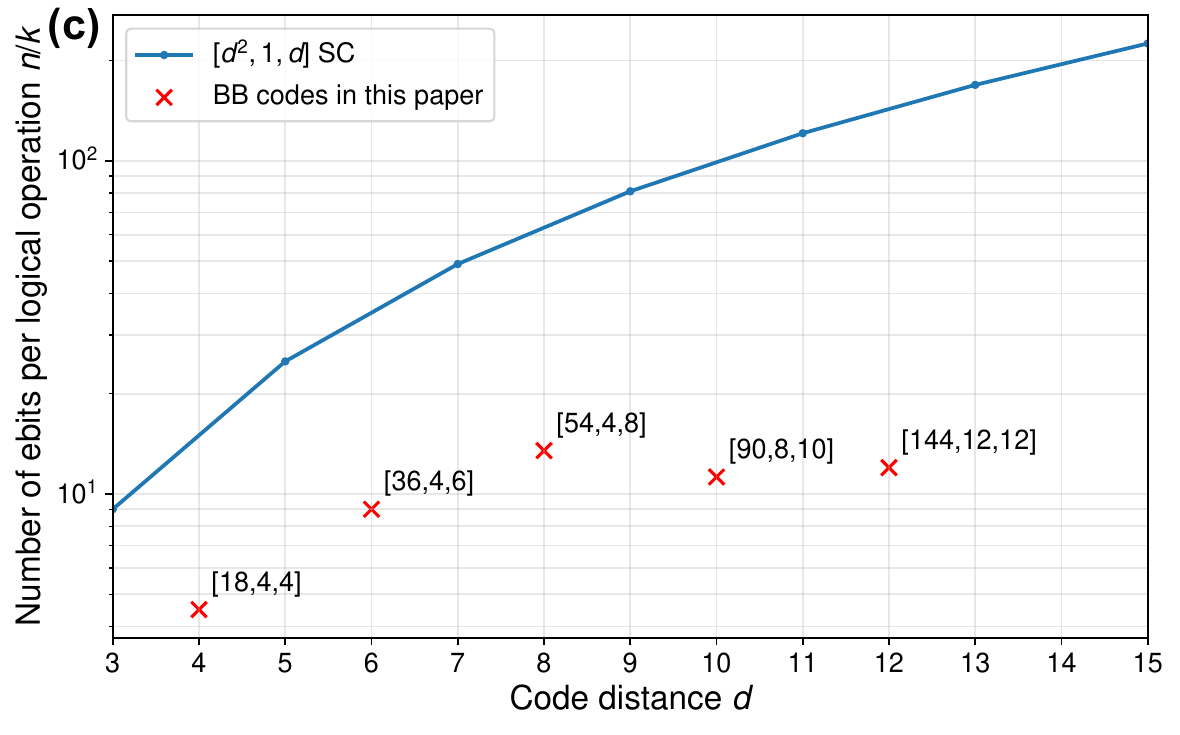}
  \end{minipage}

  \caption{\textbf{Comparing surface code lattice surgery to BB code
  transversal operations.} \textbf{(a)} Diagram showing the implementation of a
  weight 4 Pauli string using transversal CNOTs over 3 nodes. For diagrammatic brevity, non-local CNOTs have been compressed compared to Fig.~\ref{fig:arch}(b) \textbf{(b)} LER
  against Pauli string weight (equivalently number of lattice surgery
  codeblocks used in Pauli product measurement). We chose
  $p=p_{ebit}=5\times 10^{-4}$ to keep the medium distance codes
  significantly below break-even. Solid curves come from a model
  defined by Eq.E1 in the Supplementary Information of~\cite{Hugo},
  with the LER adjusted from per-round to the LER across the whole
  PPM. Shading represents number of ebit seams in merged lattice
  surgery codeblocks, which is ultimately negligible. Dashed gray and charcoal lines
  indicate results extrapolated from this paper. Purple dashed lines
  represent our circuit-level transversal simulations. \textbf{(c)} Number of
  ebits required per logical operation against code distance $d$. Blue
  line represents surface codes, plotted crosses represent the BB
  codes discussed in this paper.}

\label{fig:transversal}
\end{figure*}

\subsection{Comparing distributed transversal operations to distributed lattice surgery}
\label{sec:dls-comparison}

A seemingly intuitive way to compare transversal and lattice-surgery
approaches is to benchmark their respective implementations of a
logical CNOT.  However, this can be misleading: Lattice-surgery--based
compilation does not typically use CNOTs as a default
operation. Instead, Pauli-based compilers compile to Pauli product
measurements (PPMs), which when combined with magic-state
injection can be a universal gateset~\cite{Litinski2019,pbcAngLi}.

A weight-$w$ PPM measures an operator (Pauli String) of the form
\begin{equation}
P_1\otimes P_2\otimes \cdots \otimes P_w,
\qquad P_i\in\{X,Y,Z\},
\label{eq:ppm_def}
\end{equation}
across $w$ logical qubits.  In lattice surgery, such measurements are
implemented by temporarily merging $w$ patches (or, in a distributed
setting, creating a joint interface across modules) and repeating the
merged-code syndrome extraction for $O(d)$ rounds before
unmerging~\cite{Litinski2019}. This is the compute step, which is
typically followed by the application of a single qubit rotation
operator via magic state injection to implement an arbitrary Pauli
rotation, followed by unmerging (uncompute).

Many quantum simulation algorithms are natively expressed in terms of
repeated Pauli-string operations~\cite{paulStringRef, nielsen, HHL,
  vqeEtc, qsp}, and several important non-simulation algorithms admit
efficient PPM-based compilations~\cite{gidney2025factor2048bitrsa,
  pbcAngLi}.  Consequently, a fair comparison between transversal and
distributed lattice surgery (DLS) in a modular setting should focus on
an operation that appears natively in algorithms we care about.  We
therefore compare transversal and DLS implementations of Pauli string
compute steps.

Transversal compute/uncompute steps can be realized in several ways.
For simplicity, we focus on strings where $P_i=Z$ for all $i$, since
general Pauli strings can be reduced to this case by single qubit
Clifford gates on the participating qubits.  A direct implementation
computes the parity of $w$ data qubits onto an ancilla using $w$ CNOTs
(data as controls, ancilla as target)~\cite{paulStringRef}.  In a
distributed setting this can be non-optimal: Depending on where the
supports of the string reside relative to the ancilla (and relative to
the inter-module boundary), it may require many non-local CNOTs.
Instead, we simulate a \emph{chained} implementation~\cite{cnotChain}
that computes the same parity:
\begin{equation}
\mathrm{CX}_{1\rightarrow 2},\;
\mathrm{CX}_{2\rightarrow 3},\;\ldots,\;
\mathrm{CX}_{(w-1)\rightarrow w},\;
\mathrm{CX}_{w\rightarrow a},
\label{eq:chain-compute}
\end{equation}
where $a$ is the ancilla.  With an appropriate ordering of the
participating qubits, this construction can minimize the number of
times the chain crosses the inter-module boundary, thereby reducing
the required number of non-local CNOTs relative to the fan-in.

We evaluate the distributed transversal compute/uncompute primitive in
two ways.  First, we form an analytic/extrapolated
estimate using the non-local CNOT logical error rates reported earlier
in this paper. Second, for representative
instances we perform circuit-level simulations of Eq.~\ref{eq:chain-compute}
for qLDPC codeblocks, see Methods for details. The results are
summarized in Fig.~\ref{fig:transversal}.

Starting from the extrapolated estimates, we find that the LER of
transversal compute is an order of magnitude lower than DLS-based
compute. However, it is important to note that the equation we use to
estimate the lattice surgery LER is from~\cite{Hugo}, where they used
the MWPM decoder. We find that under the Tesseract decoder lattice
surgery $M_{ZZ}$ operations have a significantly lower LER and
approach transversal methods (Supplementary Note 6). This means
that, adjusting for differences in decoders, neither method provides a
conclusive advantage in LER.

However, in our BB instances, a single transversal non-local CNOT acts
on $k$ pairs of logical qubits in parallel.  Accordingly, the
transversal compute/uncompute step in Eq.~\ref{eq:chain-compute} corresponds
to $k$ parallel compute steps for $k$ Pauli strings across the
participating codeblocks. Recent work shows early indications that
this parallelism can be efficiently exploited in quantum simulation
algorithms~\cite{xu2025batchedhighratelogicaloperations}. To provide a
fair comparison, we should compare to a surface-code DLS implementation that
repeats the PPM $k$ times, and we report this $k$-scaled baseline in
Fig.\ref{fig:transversal}(b). Accounting for this, non-local
transversal operations on BB codes begin to show a significant advantage over
distributed lattice surgery on surface codes.

In addition, beyond LER, the resource scaling differs markedly in a
distributed setting: Performing $k$ compute steps via surface-code DLS
requires $O(n\times k)$ ebit consumptions in total across the $k$
repetitions (per seam), whereas the qLDPC transversal construction
requires only $n$ ebits to enact the $k$ steps in parallel, or $n/k$
ebits per Pauli string (per seam).  Assuming the Pauli string supports
can be mapped and rearranged efficiently across the qLDPC blocks, this
reduction in ebit consumption directly relaxes network bandwidth
demands (see Fig.~\ref{fig:transversal}c) and can mitigate
waiting-time decoherence. Moreover, in general transversal operations
are thought to be more efficient and faster to implement on the
devices that support them. This is because recent work shows
transversal operations need only O(1) syndrome extraction rounds
before/after each operation~\cite{quEraTransversalDecoding,
  quEraTransversalDecoding2} rather than O(d) required for lattice
surgery.

Our circuit-level simulations give more complex results. As the weight
increases, LER increases much more rapidly than the extrapolated
results. This is because the decoding volume of the multiple
codeblocks begins to grow exponentially and overwhelm the
decoder~\cite{duke}. This is not a fundamental limitation. Correlated,
windowed and ``ghost decoding'' approaches have been proposed and
tested on surface codes to prevent
this~\cite{quEraTransversalDecoding, quEraTransversalDecoding2,
  ghost}, with~\cite{quEraTransversalDecoding2} highlighting that
efficient decoding in this way is possible for qLDPC codes. However,
we do not believe decoders of this type are publicly available for
qLDPC codes.

Platforms with native nearest-neighbor interactions and limited
long-range connectivity will naturally favor DLS-style PPMs. But on
platforms that can efficiently implement qLDPC codes, transversal
operations and couple shared ebits transversally will be able to
exploit a significant advantage. 
\section{Discussion}
\label{sec:conc}

Our results provide a comprehensive circuit-level analysis of
fault-tolerant primitives for distributed quantum computing and their
use in Pauli string rotations, establishing key principles for
architecture design, code selection, and compiler strategies. By
moving beyond memory experiments to simulate full transversal
non-local CNOT and teleportation operations, we bridge the gap between
abstract code properties and the practical constraints of noisy,
interconnected quantum modules.

We have demonstrated that the transversal non-local CNOT consistently
achieves logical error rates up to an order of magnitude lower than
logical teleportation at the same code distance and noise levels. This
performance advantage stems from its simpler circuit structure that
uses only two codeblocks rather than three, reducing the decoding
volume. The circuit has fewer ancillary operations, measurements, and
classical feed-forward steps, thereby reducing the number of potential
error pathways. The magnitude of the difference in LER is significant,
and we found that gate-count based proxies are insufficient for
predicting the LER.

Consequently, compiler strategies should prioritize scheduling
computations across nodes using non-local CNOTs rather than frequently
moving logical state via teleportation, reserving the latter for
scenarios where it eliminates a large number of non-local gates (e.g.,
on the order of 10).

A critical finding for hardware development is the tolerance of both
primitives to relatively noisy entanglement. Our simulations show that
even when the ebit error rate ($p_{\text{ebit}}$) is an order of
magnitude worse than the PER ($p$), the primitives scale: Lower PER
and higher distances lead to lower LER. Extrapolating from our
simulations, we find that with large enough code distances ($d \sim
29$) a non-local CNOT at $p=10^{-3}$ with $p_{\text{ebit}}=10^{-2}$
can be executed with LER low enough for execution of large scale
algorithms such as Shor's algorithm. At $p=10^{-4}$ with $p_{\text{ebit}}=10^{-3}$, $d \sim11$ is sufficient.

This significantly relaxes the fidelity requirements for DQC,
suggesting that moderate-infidelity ebits (${\sim}10^{-3}-10^{-2}$)
may be sufficient for fault-tolerant applications without immediate
need for resource-intensive distillation.

Regarding the decoherence accrued while the $n$ ebits necessary for
our primitives are generated, we find that the hardware requirements
are stringent, especially as code sizes increase.  The serial
(one-at-a-time) regime requires very fast ebit generation compared to
decoherence time ($\leq 10^{-4}$ for large codes).  Importantly,
enabling $\mathcal{O}(d)$-way parallel production reduces the required
generation rates by approximately an order of magnitude
($\leq 10^{-3}$ for large codes) and can therefore be as important as
improving the intrinsic ebit fidelity itself.  The required generation
rates are within experimental feasibility for neutral atom and trapped
ion platforms, especially for the parallel generation regime.  But it
seems unlikely that either of these regimes would be suitable for
superconducting devices or other devices with decoherence times on the
order of microseconds, with ebit generation rates needing to be faster
than nanoseconds, even for smaller codes.  We hope our establishing of
these thresholds motivates further experimental development in
high-bandwidth entanglement generation.

Our comparison of surface codes and bivariate-bicycle codes presents a positive picture for the use of bivariate-bicycle codes (and
similarly performing qLDPC codes) in distributed quantum
computing. For the non-local CNOT primitive, BB codes achieve very
similar LERs to SCs. Assuming that BB codes (as well as transversal
operations) are realizable in the target hardware, then parallel
transversal CNOTs on multiple logical qubits can be a powerful and lower-cost operation compared to lattice surgery on surface codes.

This leads us to our analysis of the use of our primitives for Pauli
rotations. We simulated, via extrapolation and circuit-level modeling,
the implementation of the compute step of Pauli rotations via
transversal operations on BB codes and Pauli Product Measurements
(PPMs) via distributed lattice surgery (DLS) on surface codes. We find
that qLDPC codes are highly advantageous in a distributed
setting. Parallel transversal CNOTs on the $k$ logical qubits of BB
code blocks enable $k$ Pauli rotation compute steps to be implemented
with approximately the same LER as the DLS approach requires for a
single compute step. Furthermore, each Pauli string uses only $n/k$
ebits per operation, compared to the $n$ required per operation for
lattice surgery, allowing transversal operations to be executed with
significantly less ebits.

However, we qualify that these advantages are heavily dependent on the
underlying physical architecture. qLDPC transversal primitives only
have an advantage if they can be efficiently realized. Ideally, a
device must have non-local connectivity and be able to transversally
interact ebits and data qubits. Architectures relying on fixed,
nearest-neighbor topologies will face significant routing overheads
that negate these benefits. Notably, surface code lattice surgery has
an advantage over transversal approaches in terms of resistance to
decoherence noise, as it only needs O(d) ebits at a time. Therefore, while transversal non-local CNOTs excel on
flexible architectures, they are not necessarily competitive
or even possible on local platforms with fast decoherence times.

Looking forward, several avenues warrant further investigation. First,
our library, TMCBS, enables the study of more complex distributed
circuits and different noise models, such as simulating more complex
gadgets and/or fine grained entanglement generation methods.

Second, while we utilized Tesseract, a state-of-the-art decoder, we
observed that it struggled to decode larger circuits. Adding more code
blocks decreased the LER far more than an extrapolation of the
two-code-block CNOT case would suggest. Because there is no
fundamental theoretical limit dictating this drop in performance, we
attribute this to the decoder struggling to decode the correlated
noise spread across multiple codeblocks. Similar issues, and improved
decoders tailored to this issue, have been noted in recent literature
regarding correlated noise in transversal operations on surface
codes~\cite{ghost, duke,quEraTransversalDecoding,
quEraTransversalDecoding2}. We hope our circuit-level identification
of this bottleneck motivates the development of such decoders. In
addition, exploring robust memory protocols to protect early ebits
while the remaining $n$ ebits are produced, such as proposed in early
studies~\cite{harvard}, may be useful to realizing distributed quantum
computing.

Finally, we hope that our results and library can provide guidance for
extending experimental realization of these transversal protocols from
the single device case~\cite{quantinuum, transversalTele} on to
emerging multi-module quantum processors. This will be crucial for
refining these models and solidifying the architectural blueprint for
scalable distributed quantum computing.
\section{Methods}
\label{sec:design}

\subsection{Simulation}
\label{sec:noise}

We use STIM~\cite{stim}, a fast stabilizer simulator. For BB codes,
our noise model and syndrome extraction cycle is identical to that
of~\cite{bravyi}. For the Surface Code, our noise model and syndrome
extraction circuit is identical to those contained in~\cite{stim}. For
both code families, every time a qubit on a different device is
initialized (after measurement or in the beginning), it suffers bit
flip noise $p$. Every 1 or 2-qubit gate has 1 or 2-qubit depolarizing
noise $p$ applied. Measurement error is simulated by applying bit flip
noise $p$ to the qubit (or ebit) being measured immediately prior to
its measurement. The BB code syndrome extraction circuit features idle
noise and, in that case, there is idle error in the form of
depolarizing noise following the noise model in~\cite{bravyi}. In
addition, any ebit-data qubit interaction has the same noise as
between two data qubits. Ebit creation noise is modeled as 2-qubit
depolarizing noise applied to the ebits after they have been perfectly
initialized, as in~\cite{UCR, Hugo}.

The code used to generate STIM circuits and decode them was initially
inspired by the repository associated
with~\cite{gong2024lowlatencyiterativedecodingqldpc}, which generates
single-codeblock BB or surface-code memory experiments and decodes
them (e.g., with BP-OSD~\cite{bposd}). Building on this starting
point, we introduced new classes, functions, and data structures that
enable generation of STIM circuits for logical operations across an
arbitrary number of code blocks using local gates and non-local gates
via explicit ebits, including support for mid-logical-circuit
measurements, classically-conditioned operations, and tracking of
multiple code-block observables.

While codes/libraries developed to evaluate lattice surgery or
transversal operations in the literature may be easily modified to
ensure that certain physical operations have noisier rates in an
attempt to mimic noisy ebits, this does not capture the full effect of
non-local operations. The additional measurements, syndrome extraction
rounds, ancillas, gates and actual physical ebits introduce
significant error sources and significantly increase the complexity of
error propagation, which also increases the challenges of
decoding. Simulating fully non-local operations requires the creation
of circuit-level simulations between multiple code blocks with actual
physical ebits. As far as we are aware, there are no pre-existing
libraries that allow the simulation of arbitrary transversal
operations between arbitrary code blocks. This motivated the
development of our library TMCBS.

In addition, our library supports scalable execution via MPI
parallelism, allowing us to run on hundreds to $>10^3$ CPU cores on
supercomputers and obtain sufficient statistics at low PER/LER for
large logical circuits using modern decoders such as Tesseract.

\subsection{Transversality}

A transversal gate in QEC is a logical operation that can be
implemented by applying the corresponding physical gate to each data
qubit in a code block. Logical operations are those that preserve
distance and the code space. For a code with $n$-qubit logical states,
applying a physical operation $U$ to each data qubit acts as
$U^{\otimes n}$(single-block), or for paired interactions such as a
CNOT, $U_{\text{CNOT}}^{\otimes n}$ between blocks. Transversal
operations do not propagate errors within a code block and are fault
tolerant, as they preserve the code space.

By the Eastin-Knill theorem~\cite{eastin}, no QEC code can perform
{\em universal} fault tolerant quantum computation using only
transversal gates. All CSS codes have directly transversal
CNOT~\cite{transCNOT}. Both Rotated surface codes and BB codes have
directly transversal Pauli gates. 

Implementing syndrome extraction rounds after applying Hadamard gates
typically requires fold-transversal methods for self-dual codes and
automorphism-based methods for non-self dual codes~\cite{ftam}. But we
only apply Hadamard gates transversally in the teleportation primitive
(i.e., we use it on the ebits for both primitives but this is a
physical operation, where the ebits are not encoded). In this
primitive, we apply it to CB2 and then CB1. But we do not perform any
syndrome extraction rounds on either of those two codeblocks after the
application of the first Hadamard, before they are
measured. Therefore, we do not need or use fold transversal or
automorphism methods.

\subsection{Subthreshold scaling}

The usual deep-subthreshold scaling form is defined as
\begin{equation}
P_{L}(p,d)\;\approx\;\alpha\!\left(\frac{p}{p_{\mathrm{th}}}\right)^{\!(d+1)/2},
\label{eq:scaling}
\end{equation}
which encodes the expected exponential-in-distance suppression below
threshold. At a fixed $p<p_{\mathrm{th}}$, define
$r=p/p_{\mathrm{th}}<1$; then increasing $d$ by two multiplies $P_L$
by $r$. Using any reference data point $(d_0,P_0)$ at the same $p$,
the distance needed to reach a target $P_\star$ follows
\begin{equation}
d \;=\; d_0 \;+\; 2\,\frac{\ln(P_\star/P_0)}{\ln r}.
\label{eq:distance_formula}
\end{equation}

\subsection{Non-local CNOT circuit}

On QC1, a physical CNOT is performed between every data qubit $k$ in
CB1 and ebit $k$. Then, every ebit on QC1 is measured (in our case in
the Z-basis). The parity of each measurement (0 or 1) is used next to
conditionally apply an X gate to ebit $k$ located on QC2. A CNOT is
subsequently performed between ebit $k$ and data qubit $k$ in CB2. A
Hadamard gate is applied to every ebit on QC2 before every ebit on QC2
is then measured (in our case, again, in the Z-basis). Finally, the
parity of the measurement determines whether or not a Z gate is
conditionally applied on CB2. The above combination of operations
results in a FT CNOT between CB1 and CB2, with each logical qubit on
CB1 being a control for its corresponding target logical qubit on
CB2. The circuit is identical (other than differences due to the
number of qubits or in the syndrome extraction rounds, and the transversal
operations applied are identical) between the surface codes and BB
codes.

For simulation purposes, we noisily initialize the two code blocks,
apply 4 syndrome extraction cycles to each, perform the non-local CNOT
between the two code blocks and then perform 3 syndrome extraction
cycles on each before measuring both code blocks in the Z basis. As is
common in other works, STIM detectors are only applied to Z
stabilizers

as there should be little difference between
X and Z logical error rates.

\subsection{State teleportation circuit}
\label{sec:noise}

Using a non-local CNOT, we can create a logical Bell state(s) between
two code blocks located on different devices. Using these, we can
teleport a codeblock using a similar procedure to the teleportation
procedure in~\cite{quantinuum}. The circuit we use is shown in
Fig.1(c).  Code blocks CB1, CB2 and CB3 are noisily
initialized before 4 syndrome extraction cycles are applied to each
code block. Next, the logical qubits of CB2 and CB3 are entangled into
logical Bell states using a transversal Hadamard on CB2 and a
non-local CNOT between CB2 and CB3.

We then perform a logical Bell state measurement on CB1, with
corrections made to CB3. Finally, we perform 3 rounds of syndrome
extraction on CB3 before measuring in the Z-basis.  The circuit is,
again, identical (other than differences due to the number of qubits
or the syndrome extraction rounds, the transversal operations applied
are identical) between surface codes and BB codes.

\subsection{Decoherence}
\label{sec:decoMethods}

To model decoherence noise, we use equations 9 and 10
from~\cite{PhysRevA.90.062320}, repeated below:

$$p_X = p_Y = \frac{1-e^{-t/T_1}}{4}$$
$$p_Z = \frac{1-e^{-t/T_2}}{2} - \frac{1-e^{-t/T_1}}{4}$$.

These are applied to each ebit on both QC1 and QC2 via the STIM
function \texttt{PAULI\_CHANNEL\_1} with noise tuple $(p_x,p_y,p_z)$.

We set $T_1=T_2$. $t$ is defined as follows: Say we are producing $n$
ebits for an operation between two $[n,k,d]$ codeblocks and each ebit
takes $t_{gen}$ to produce. Then, immediately after the $n$th ebit has
been produced, the $j$th ebit to be generated has been de-cohering for
$t=(n- j)t_{gen}$. In a device where $O(d)$ ebits can be produced
simultaneously, we assume $f=d$ ebits can be generated
simultaneously for surface codes. For BB codes, we define $f$ as the
closest integer to $d$ where $f$ divides $n$. Ebits in the $l$th line
have therefore been decohering for $t=(n- j)t_{gen}$.

We do not consider the stochasticity of ebit generation in this
paper. This is because we find that the effect it has on the age of
the oldest ebit(s) is relatively negligible compared to the range of
parameters we consider in this work, see Supplementary Note 5 for
more detail. Detailed consideration of ebit generation is left to
future work.

\subsection{Pauli string compute}

For the circuit level simulations, we simulate the following: for a
Pauli string of weight $w$, we use $w+1$ codeblocks, the final
codeblock being the ancilla. We noisily initialize the codeblocks in
the zero state and then apply four syndrome extraction rounds to every
codeblock. We apply a chain of CNOTs to the codeblocks following
Eq.~\ref{eq:chain-compute}. The CNOTs are either local (0) or non-local CNOTs
(1) and we obey the following pattern as $w$ increases: $w=2$: 1,
$w=3$: 10, $w=4$: 101, $w=5$: 1010, $w=6$: 10101. As an example,
Fig.~\ref{fig:transversal}(a) shows $w=4$. Local CNOTs are transversal
CNOTs with noise on each physical CNOT equal to PER, as is the case in
other experiments in this paper. Non-local CNOTs are carried out using
the circuit in Fig.~\ref{fig:arch}(c), where ebits experience a
noise level equal to the PER. After each CNOT between codeblocks $m$
and $m+1$, we carry out a single syndrome extraction round solely on
codeblocks $m$ and $m+1$. Once we have applied the chain of CNOTs to
the codeblocks in question, we apply a local CNOT from the final
codeblock to the ancilla. We then noisily measure all codeblocks,
including the ancilla. We report the LER from these experiments in
Fig.~\ref{fig:transversal}(b).

We use the following equation for the extrapolated transversal
results: If a non-local CNOT has LER $l$ then for
Fig.~\ref{fig:transversal}(b) $y=1-(1-l)^x$. For the
$\llbracket 54,4,8\rrbracket$ code we obtained $l$ by running a
non-local CNOT simulation at $p=p_{ebit}=5\times10^{-4}$ giving LER
$2.00\times10^{-6}$ (CI: $[8.73\times10^{-7},
3.84\times10^{-6}]$). For the $\llbracket 144,12,12\rrbracket$ code we
had trouble simulating a single non-local CNOT at this PER. This can
be seen in the size of the error bar (2 errors on $4.32\times10^{8}$
shots) for simulations with a slightly higher LER than that operation,
namely a Pauli string weight of 2, which represents a non-local CNOT
and a local CNOT across 3 codeblocks. Therefore, we instead
extrapolated from the $\llbracket 121,1,11\rrbracket$ result in
Fig.\ref{fig:all}(c) at $p=5\times10^{-4}, p_{ebit}=10^{-3}$ (which
has the same LER as $p_{ebit}=10^{-4}$).

For the lattice surgery extrapolations, we use Equation~E1
from~\cite{Hugo}, which gives the $X$-logical error rate per syndrome
extraction round, $P_L^X$, for a distributed rotated surface code patch of
size $d_x \times d_z$ with $n_{\text{seam}}$ seams. In our setting, the
$X$-distance $d_x = d$ is the code distance of the surface code patches
being merged, and the $Z$-distance $d_z = w \cdot d$ is the total length
of the merged routing patch, where $w$ is the Pauli string weight
(equivalently, the number of lattice surgery codeblocks). The number of
seams $n_{\text{seam}}$ equals the number of inter-processor boundaries
crossed by the merged patch. Since each PPM
requires $d$ syndrome extraction rounds, we convert the per-round rate to a
per-PPM rate via $LER_{PPM} = 1-(1-P_L^X)^{d}$. To report the LER for $N$
PPM operations, we compose as
$LER_{N\times PPM} = 1-(1-LER_{PPM})^{N}$.

\subsection{Decoding}

A detector error model (DEM) is then extracted from the STIM
circuit. A new parity-check matrix is derived from the DEM, where each
``stabilizer'' (row) corresponds to a detector. Each ``data qubit''
(column) now represents a distinct error mechanism that flips one or
more detectors and possibly an observable. In the case of just the
non-local CNOT circuit, there are $2k$ logical observables for a
$\llbracket n,k,d\rrbracket$ code block. Each logical observable is
the Z basis measurement of a logical qubit in one of the two code
blocks. In the case of the teleportation circuit, there are $k$
logical observables. Each logical observable is the Z basis
measurement of the logical qubits of code block 3 as defined in
Fig.~\ref{fig:arch}(c). The syndrome in both cases is a vector, where
each entry represents whether a detector is 0 or 1 after the circuit
has been noisily executed. This PCM and syndromes are then decoded
using the Tesseract decoder~\cite{teser} under default settings.

Note that logical error rate in this work refers to the logical error
rate of the entire operation, not per-round or per-logical-qubit or
per-codeblock. For example, a LER of $10^{-5}$ for a non-local CNOT
means that there is a $10^{-5}$ probability that at least one logical
error has occurred on any logical qubit or code block during the
execution of the entire algorithm and the measurement of the logical
qubits after. This is the same for the Pauli string simulations, LER
refers to the error that any error occurred on any logical qubit or
codeblock during the execution of the compute step and the measurement
of the logical qubits. Whilst per-round LERs are reported in memory
experiments, this metric makes less sense in the context of logical
operations across multiple codeblocks.

\subsection{Statistics}
All error bars in figures indicate the region where the Bayes' factor
is $\leq 1,000$. Assuming that logical errors are binomial
distributed, the regions indicated by the error bars are those where
the probability of the LER being in that region is no more than
$1,000$ times less than the best hypothesis, i.e., the LER is the
number of logical errors divided by the number of shots. The error
bars have the same meaning for all subsequent figures. If a PER lacks
data points, the omission is attributed to the wall clock time being too
small to generate a statistically significant amount of errors at this
level.

\newline
\newline
{\setlength{\parindent}{0cm} \textbf{Acknowledgments:} We thank Joschka Roffe for useful discussions and feedback. This work was
  supported in part by NSF awards MPS-2531350,
OSI-2410675, PHY-2325080, CISE-2316201, MPS-2120757, CCF-2217020, and
PHY-1818914 as well as DOE DE-SC0025384. This research used resources of the National
Energy Research Scientific Computing Center, a DOE
Office of Science User Facility supported by the Office of
Science of the U.S. Department of Energy under Con-
tract No. DE-AC02-05CH11231 using NERSC award
ASCR-ERCAP0037552.}\newline

{\setlength{\parindent}{0cm}\textbf{Author contributions:}
  J.S. conceived the study, developed the TMCBS library, performed the simulations, analyzed the data and wrote the initial draft of the paper. M.W. provided guidance regarding QEC codes and decoding and provided support with writing TMCBS, they also contributed to the QEC section in the supplementary info. F.M. supervised the project and guided design as well as execution of the study. All authors discussed the results and contributed to the writing of the manuscript.}
\newline

{\setlength{\parindent}{0cm}\textbf{Code availability:} We will publish TMCBS upon publication.}

\bibliography{sample-base}

\end{document}


\newcommand{\innerProduct}[2]{\langle #1 \vert  #2 \rangle} 
\newcommand{\memk}[1]{\ket{\boldsymbol{#1}}}
\newcommand{\memb}[1]{\bra{\boldsymbol{#1}}}
\renewcommand{\topfraction}{.85}
\renewcommand{\bottomfraction}{.85}
\renewcommand{\floatpagefraction}{.95}
\renewcommand{\dblfloatpagefraction}{0.95}
\renewcommand{\textfraction}{0.07}
\renewcommand\thesection{\arabic{section}}

\setcounter{topnumber}{2}
\setcounter{dbltopnumber}{2}
\renewcommand{\topfraction}{0.95}
\renewcommand{\dbltopfraction}{0.95}
\renewcommand{\textfraction}{0.05}
\renewcommand{\floatpagefraction}{0.9}
\renewcommand{\dblfloatpagefraction}{0.9}

\renewcommand\thesection{Supplementary Note~\arabic{section}}

\renewcommand{\cftsecpresnum}{}
\renewcommand{\cftsecaftersnum}{:\quad}
\setlength{\cftsecindent}{0pt}
\settowidth{\cftsecnumwidth}{Supplementary Note 10: \quad}

\makeatletter
\renewcommand{\thefigure}{\arabic{figure}}
\renewcommand{\thetable}{\arabic{table}}
\def\fnum@figure{Supplementary Figure~\thefigure}
\def\fnum@table{Supplementary Table~\thetable}

\makeatother

\title{Supplementary Information for \emph{Transversal fault tolerant distributed quantum computing operations}}

\author{John Stack}
\email{Contact author: jstack@ncsu.edu}
\author{Ming Wang}
\author{Frank Mueller}
\affiliation{Department of Computer Science, North Carolina State University, Raleigh, North Carolina, USA.}

\begin{abstract}
This document contains supplementary information to the article:  \emph{Transversal fault tolerant distributed quantum computing operations}.
\end{abstract}
\maketitle

\makeatletter\@starttoc{toc}\makeatother

\section{Quantum Error Correction (QEC)}

Let $\mathcal{P}=\langle X,Y,Z \rangle$ denote the single-qubit Pauli
group. The $n-$qubit Pauli group is then defined as
$\mathcal{P}_n=\{ P_1\otimes P_2 \otimes \cdots\otimes P_n | P_i\in
\mathcal{P}\}$. An $\llbracket n,k,d\rrbracket $ stabilizer code,
defined by the stabilizer group $\mathcal{S} \subset \mathcal{P}_n$,
encodes $k$ qubits of logical information into a $n$-qubit physical
qubit block and can correct up to $\lfloor (d-1)/2 \rfloor$ errors,
with an encoding rate of $k/n$. The code space $\mathcal{C}$ is the
common $+1$ eigenspace of $\mathcal{S}$ given by
\begin{equation}
    \mathcal{C}=\{\ket{\psi} \:|\: s\ket \psi=+\ket{\psi}\}. 
\end{equation}

To construct a valid stabilizer code, the stabilizer group
$\mathcal{S}$ must be an Abelian subgroup of the $n$-qubit Pauli group
$\mathcal{P}_n$ such that $-I \notin \mathcal{S}$. For qubit
stabilizer codes, each generator $g \in \mathcal{S}$ can be written as
a tensor product $g = P_1 \otimes \cdots \otimes P_n$, where
$P_i \in \{I, X, Y, Z\}$. Each generator can be mapped to a binary
vector $\mathbf{g} = [\mathbf{g}^X \mid \mathbf{g}^Z]$ of length $2n$
using the symplectic representation. The mapping from each
single-qubit Pauli operator $P_i$ to a binary pair $(g^X_i, g^Z_i)$ is
defined as
\begin{equation}
(g^X_i, g^Z_i) =
\begin{cases}
(0, 0) & \text{if } P_i = I \\
(1, 0) & \text{if } P_i = X \\
(1, 1) & \text{if } P_i = Y \\
(0, 1) & \text{if } P_i = Z.
\end{cases}
\end{equation}
Collecting these binary vectors as rows yields the stabilizer check matrix
\begin{equation}
H = \left[\begin{array}{c|c}
H_X & H_Z
\end{array}\right] \in \mathbb{F}_2^{(n-k) \times 2n},
\end{equation}
where $H_X, H_Z \in \mathbb{F}_2^{(n-k) \times n}$ correspond to the
$X$ and $Z$ components of the stabilizer generators in binary form.

Calderbank-Steane-Shor (CSS) codes form a special subclass of
stabilizer codes in which the stabilizer group $\mathcal{S}$ can be
decomposed into two disjoint subsets containing only $X$-type and
$Z$-type stabilizers. Consequently, the parity-check matrix of a CSS
code takes the block-diagonal form
\begin{equation}
    H=\left[
\begin{array}{c|c}
         H_X & \bm{0}\\
         \bm{0}&H_Z
    \end{array}\right].
\end{equation}
Since $\mathcal{S}$ is an Abelian group, the commutativity condition
requires that $H_XH_Z^T=\bm{0}$ must be satisfied to ensure the
validity of the CSS code.

An error that exceeds a code's correction capability can cause a
logical error. That is, after error correction, the residue error acts
as a logical operator and changes the logical state of the code. Such
errors are undetectable as they commute with all stabilizers.

\emph{Decoding:} A QEC cycle involves a syndrome
extraction round, measurement of stabilizer qubits and then correction
of inferred errors (either physically, or in a Pauli frame). Each
stabilizer qubit corresponds to a row of either $H_X$ or $H_Z$. The
result of its measurement indicates the parity of the data qubits of
that row.

Using these measurements, we can therefore produce a syndrome vector
\(\mathbf{s} \in \mathbb{F}_2^{n-k}\). Assuming that CSS codes are
used for error correction and that errors are of the form
\( E = \bigotimes_{i=1}^n P_i \), represented as a binary vector
\( \mathbf{e} = (\mathbf{e}_X | \mathbf{e}_Z) \in \mathbb{F}_2^{2n}
\), we can infer that the errors that have occurred on the data qubits
related to the syndrome $\mathbf{s}$ by the following equations:
\begin{equation}
\begin{split}
        \mathbf{s}_X = H_X\odot \mathbf{e}_Z= H_X \mathbf{e}_Z \mod 2\\
        \mathbf{s}_Z = H_Z\odot \mathbf{e}_X= H_Z \mathbf{e}_X \mod 2.
\end{split}
\end{equation}  

Given this relation, a decoder for an \(\llbracket n,k,d\rrbracket \)
stabilizer code is a classical algorithm that infers a recovery
operation \( \hat{E} \in \mathcal{P}_n \) from measured syndromes
$\mathbf{s}_X, \mathbf{s}_Z$ that removes the syndrome. The decoder
solves the inverse problem,
\( H_{\{X, Z\}} \odot \mathbf{\hat{e}}_{\{Z, X\}} = \mathbf{s}_{\{Z,
  X\}} \), to estimate \( \mathbf{\hat{e}} \). The same syndrome can
correspond to multiple errors \( \mathbf{e} \) as $H$ has fewer rows
than columns. In these cases, a decoder typically chooses the error
with the lowest Hamming weight, \( \mathbf{\hat{e}} \), i.e., the most
likely error (highest probability).

\emph{Threshold:} The threshold of a QEC code family is the physical
error rate past which increasing the code distance decreases the
logical error rate~\cite{thresh}.

\section{Surface Codes}

The rotated surface code (SC) is a topological QEC code defined on a square
lattice of \( d \times d \) data qubits (aka.  plaquette) with
parameters \(\llbracket d^2, 1, d\rrbracket \). The stabilizer group
\( \mathcal{S} \) consists of \( X \)- and \( Z \)-type operators
acting on each plaquette, with bulk stabilizers of weight four and
edge stabilizers of weight two. For a coordinate \( (i,j) \) in the
rotated lattice, the \( X \)-stabilizers \( S_X^{(i,j)} \) and
\( Z \)-stabilizers \( S_Z^{(i,j)} \) are defined as
\begin{equation}
    S_X^{(i,j)} = \bigotimes_{(k,l)\in \partial(i,j)} X_{k,l}, \quad S_Z^{(i,j)} = \bigotimes_{(k,l)\in \partial(i,j)} Z_{k,l},
\end{equation}
where \( \partial(i,j) \) denotes the qubits adjacent to plaquette
\( (i,j) \). This code is also a CSS code. Logical operators are pairs
of \( \overline{X} \) or \( \overline{Z} \) strings spanning the
lattice's diagonal with minimum weight \( d \). Under independent
circuit-level Pauli noise, the code achieves a threshold
\( p_{\text{th}} \approx 1\% \)~\cite{surfacecode}. The rotated surface code's constant
weight stabilizers and nearest-neighbor connectivity mean that it is
well suited to many hardware implementations and has been featured in
many physical experiments~\cite{google1, surfaceCode2}.

\section{qLDPC Bivariate Bicycle Codes}

Bivariate bicycle (BB) codes form a class of CSS codes defined by two
integers, $l,m$, and two bivariate polynomials: $a(x,y), b(x,y)$, over
the quotient ring $\mathbb{F}_2[x,y]/(x^l-1,y^m-1)$~\cite{bravyi}. Let
$I_i$ denote the identity matrix of size $i$, and let $S_j$ be the
cyclic permutation matrix of size $j$, defined as $S_j=I_j>>1$, where
$>>$ 1 represents a single application of a cyclic shift to the matrix
columns. We can identify the variates $x$ and $y$ as matrices
\begin{equation}
    x=S_l\otimes I_m,\quad y=I_l\otimes S_m.
\end{equation}
This representation allows the polynomials $a(x,y)$ and $b(x,y)$ to be
naturally expressed as $lm\times lm$ matrices $A$ and $B$,
respectively. Since $xy=yx$ holds due to the mixed-product property of
the Kronecker product, matrix multiplication among these
representations remains commutative. The parity-check matrices for BB
codes are then given by
\begin{equation}
    H_X = [A|B], \quad H_Z=[B^T|A^T].
\end{equation}
This construction guarantees a valid CSS code because the
commutativity of $A$ and $B$ ensures that $H_XH_Z^T=AB+BA=2AB=0$ in
$\mathbb{F}_2$, satisfying the CSS code commutativity condition. 

BB codes' main advantage over the surface code is their high encoding
as a function of code distance. Encoding rate $r = \frac{k}{n+c}$ is
the number of logical qubits encoded in a code over the number of
physical $n$ and stabilizer qubits $c$ used~\cite{bravyi}. The surface
code has encoding rate $r \approx \frac{1}{2 d^{2}} =
\frac{1}{2n}$. But BB codes have far better encoding rates. The codes
considered in this paper have encoding rates up to 10 times higher
than similar distance surface codes. Additionally, they have
low-weight stabilizers that can be fixed to degree 6~\cite{bravyi}
(unlike some other qLDPC codes) and can be efficiently implemented on
neutral atom computers~\cite{effNeutral}. The BB codes we use in this
paper are from~\cite{wang2024coprimebivariatebicyclecodes, bravyi}. Their specific parameters are in ~\ref{sec:param}. Instructions on how
to generate the code and syndrome extraction circuit, given these
parameters, are in~\cite{bravyi}.

\section{Bivariate Bicycle Code Parameters}
\label{sec:param}
The parameters for each BB code used in this paper are in
Supplementary Table~\ref{tab:bbparam}.
\begin{table}[htb]
\caption{\label{tab:bbparam} \bf{Parameters for generating the BB codes used in this paper.}   }
\begin{ruledtabular}
\begin{tabular}{lcccr}
\textrm{Code}&
\textrm{$l$}&
\textrm{$m$}&
\textrm{$a(x, y)$}&
\textrm{$b(x,y)$}\\
\colrule
$\llbracket 18,4,4\rrbracket$ \cite{wang2024coprimebivariatebicyclecodes} & 3 & 3 & $1+x+y$ & $1+x^2+y^2$  \\
$\llbracket 36,4,6\rrbracket$ \cite{wang2024coprimebivariatebicyclecodes} & 3 & 6 & $x+y^2+y^3$ & $1+x^2+y$  \\
$\llbracket 54,4,8\rrbracket$ \cite{wang2024coprimebivariatebicyclecodes} & 3 & 9 & $x+y+y^3$ & $1+x^2+y^2$  \\
$\llbracket 90,8,10\rrbracket$ \cite{bravyi} & 15 & 3 & $x^9+y+y^2$ & $1+x^2+x^7$  \\
$\llbracket 144,12,12\rrbracket$ \cite{bravyi} & 12 & 6 & $x^3+y+y^2$ & $x+x^2+y^3$ \\
\end{tabular}
\end{ruledtabular}
\end{table}

\section{Ebit Generation Stochasticity}

\label{sec:age}

\begin{figure*}[t]
  \centering
  \begin{minipage}[t]{0.49\textwidth}
    \includegraphics[width=\linewidth]{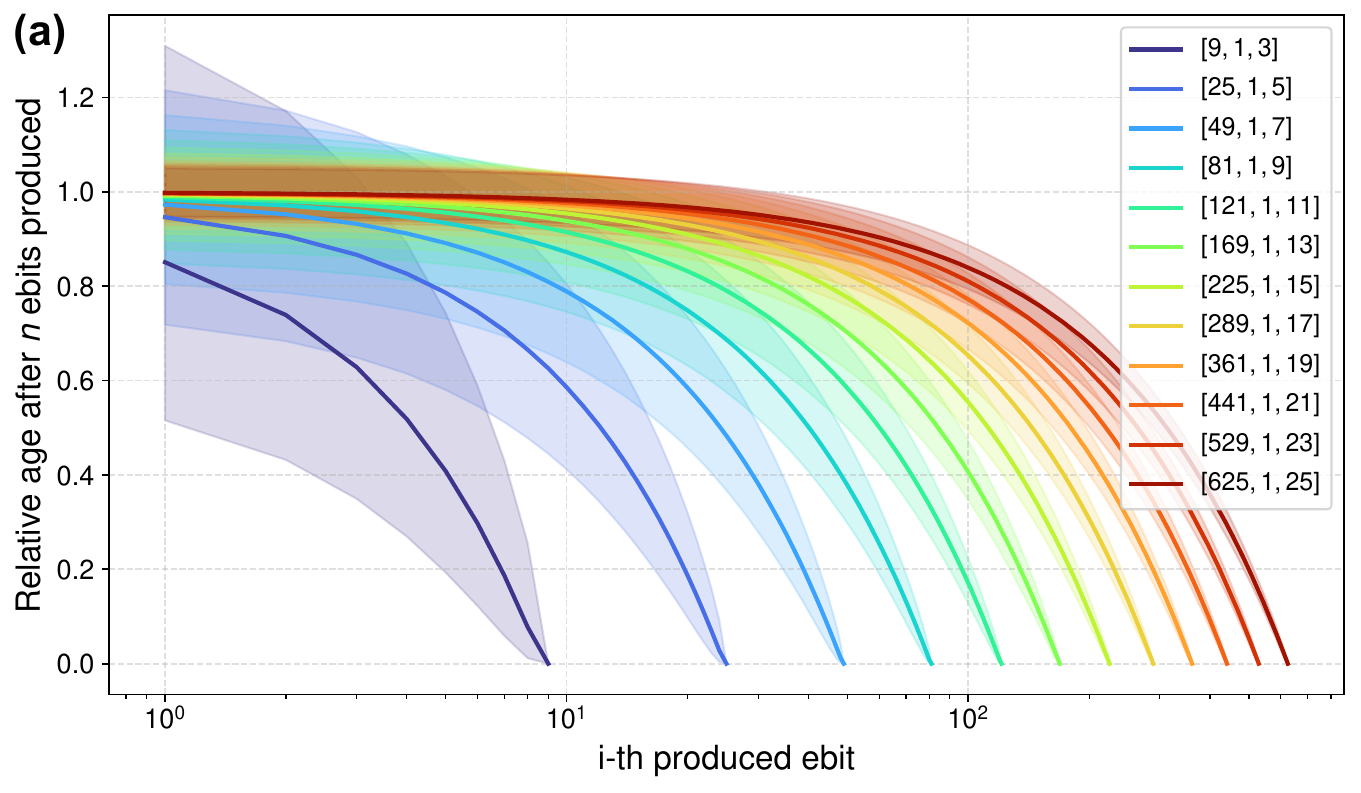}
  \end{minipage}\hfill
  \begin{minipage}[t]{0.49\textwidth}
    \includegraphics[width=\linewidth]{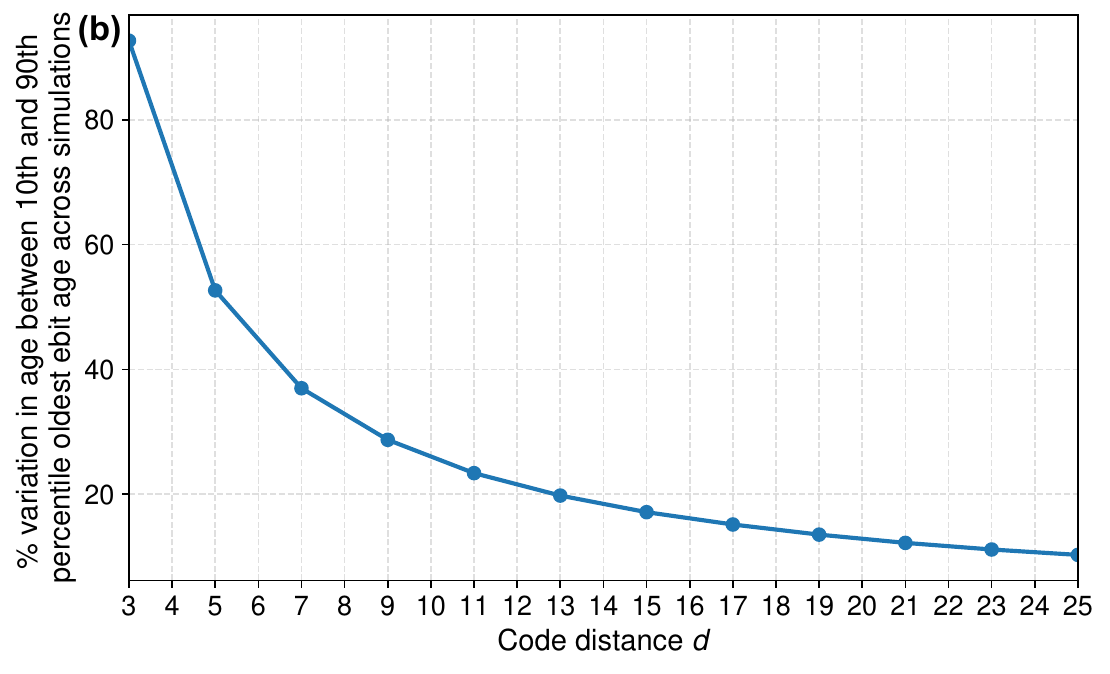}
  \end{minipage}

  \caption{\textbf{Monte Carlo simulation of the effect of
      probabilistic generation on ebit ages.} \textbf{(a)} Relative age of i-th
    produced ebit compared to the time taken to produce all $n$
    required ebits. Line is the mean time. Bands show the 10th and
    90th percentiles across $50,000$ Monte Carlo simulations. \textbf{(b)}
    Percentage variation between the 10th and 90th percentile oldest
    ebit age across $200,000$ Monte Carlo simulations.}
  \label{fig:ebit-ages}
\end{figure*}

To assess whether stochastic ebit generation significantly impacts
ebit ages compared to a deterministic model, we simulated how the age
of ebits relative to the time taken to produce the $n$ ebits required
for a transversal operation changes across Monte Carlo simulations. 
We model ebit creation as repeated attempts at rate $s$ (attempts per
second), each succeeding with probability $p$. In the continuous-time
Poisson limit, this yields a success process with rate $\lambda = sp$,
so the time to accumulate $n$ ebits satisfies
$T_n \sim \mathrm{Gamma}(n,\text{rate}=\lambda)$. The age of the
$i$-th created ebit at the moment the batch completes is
$A_i = T_n - T_i$, where $T_i$ is the creation time of the $i$-th
success. After normalizing by $\mathbb{E}[T_n]=n/\lambda$, the
distribution of $A_i/\mathbb{E}[T_n]$ becomes independent of $s$ and
$p$ and depends only on $n$ and the index $i$. Results of this
simulation for different surface codes is shown in Supplementary
Fig.~\ref{fig:ebit-ages}(a).

In Supplementary Fig.~\ref{fig:ebit-ages}(b) we concentrate on the variation in the
oldest ebit age as code distances increase. We plot the percentage
variation in age between the 10th and 90th percentile of oldest ebit
age across the simulations. Importantly, we see that whilst this
variation can be relatively high for the smallest codes, as we
approach larger codes the variation asymptotically decreases. For
codes with $d=7$ or larger this variation is unlikely to have a major
effect compared to other factors considered in this paper. When hardware
parameters are more concrete and a particular device is being
specifically targeted, then consideration of such stochasticity could
potentially be important. But at the level of generality considered in
this paper, considering stochasticity would not cause significant
changes in results and would increase the amount of compute required
to perform our simulations significantly due to the increase in the
number of shots required.

\section{$M_{ZZ}$ LER Under Different Decoders}

\label{sec:Mzz}
\begin{figure}[H]
  \centering
  \includegraphics[width=0.5\linewidth]{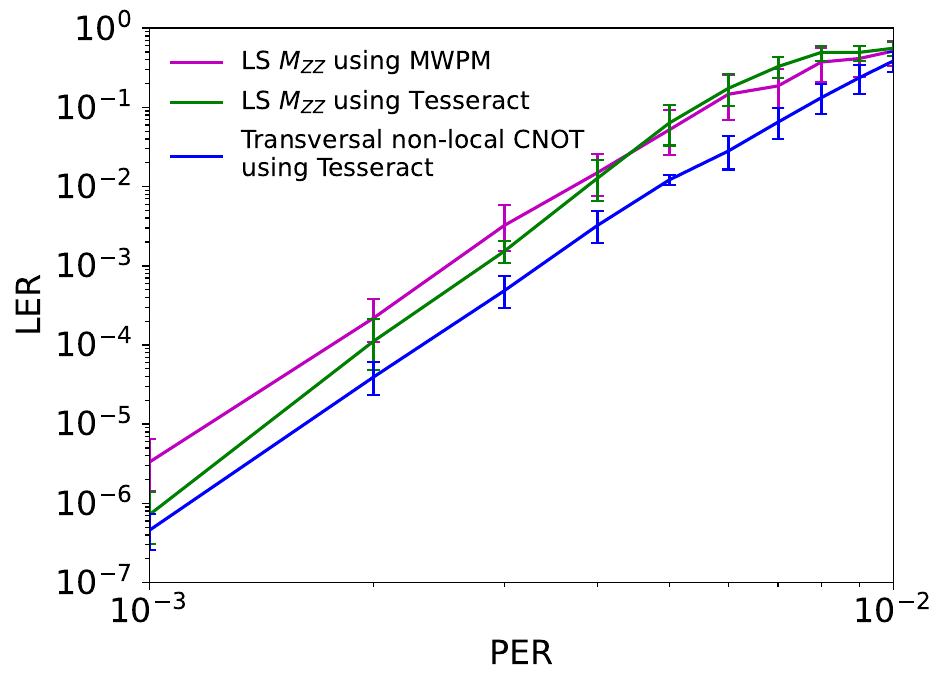}
  \caption{\textbf{Logical error rate vs physical error rate for
      different operations.} Distributed lattice surgery $M_{ZZ}$
    circuits are produced via code from~\cite{UCR} between $d=11$
    rotated surface code codeblocks using the CAT gadget setting. Ebit
    noise is set to PER. }
  \label{fig:Mzz}

\end{figure}

In the literature, lattice surgery $M_{ZZ}$ operations have been shown
to have a higher LER than transversal operations~\cite{duke} in the
non-distributed case. In Supplementary Fig.~\ref{fig:Mzz} we observe this
consistently for lattice surgery circuits decoded via MWPM. But for
distributed LS circuits decoded with the Tesseract decoder, we find
that past $p=4\times10^{-3}$, LERs begin to become competitive with
the transversal non-local CNOT operation. This may also be true in the
non-distributed case.

\section{BP-OSD Simulations}

\begin{figure}[H]
  \centering
  \includegraphics[width=0.45 \linewidth]{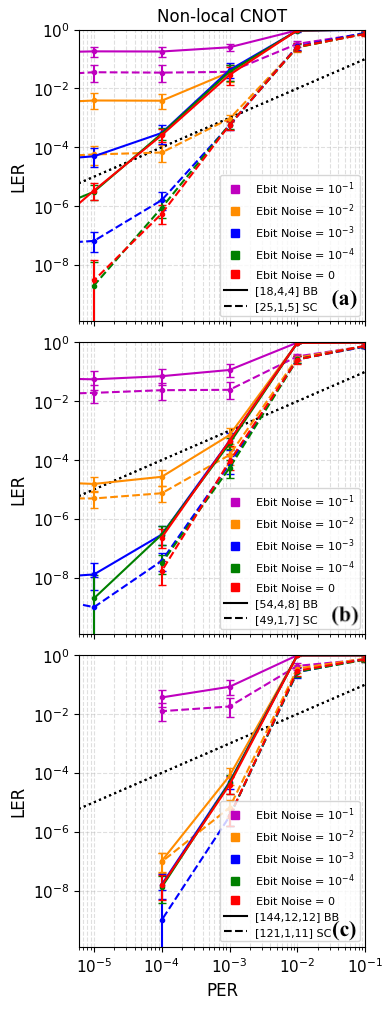}

  \caption{\textbf{Performance of distributed operations for different
      ebit error rates decoded using BP-OSD.} Physical noise rate
    against LER for different ebit noise rates for non-local CNOT
    \textbf{(a-c)} or teleportation \textbf{(d-f)} between pairs of SC or BB code
    blocks. Solid lines indicate BB codes, dashed lines indicate SC
    codes. }
  \label{fig:all}
\end{figure}

In Supplementary Fig.~\ref{fig:all}, we present the performance of
distributed operations decoded using BP-OSD~\cite{bposd} with a
maximum of $10^4$ BP iterations and $\texttt{OSD\_ORDER} = 7$. We find
that under BP-OSD (as reported in~\cite{teser} in the monolithic
case), there is a significant gap in performance between BB and SC
circuits. This gap almost disappears under Tesseract (Main Text).

\addtocontents{toc}{\protect\renewcommand{\protect\cftsecpresnum}{}}
\addtocontents{toc}{\protect\renewcommand{\protect\cftsecaftersnum}{}}
\bibliography{sample-base}